\documentclass[traditabstract]{aa}

\usepackage{txfonts}
\usepackage{graphicx}
\usepackage{epsfig,graphics,lscape,url}
\usepackage{natbib}
\bibpunct{(}{)}{;}{a}{}{,}
\usepackage{longtable}

\begin{document}

\title{Atomic data for neutron-capture elements}
\subtitle{II.  Photoionization and recombination properties of low-charge krypton ions}

\author{N.\ C.\ Sterling$^{\ref{inst1}}$\thanks{NSF Astronomy and Astrophysics Postdoctoral Fellow}}

\institute{Michigan State University, Department of Physics and Astronomy, 3248 Biomedical Physical Sciences, East Lansing, MI, 48824-2320, USA, \email{sterling@pa.msu.edu} }\label{inst1}


\abstract{We present multi-configuration Breit-Pauli distorted-wave photoionization (PI) cross sections and radiative recombination (RR) and dielectronic recombination (DR) rate coefficients for the first six krypton ions.  These were calculated with the AUTOSTRUCTURE code, using semi-relativistic radial wavefunctions in intermediate coupling.  Kr has been detected in several planetary nebulae (PNe) and H~II regions, and is a useful tracer of neutron-capture nucleosynthesis.  PI, RR, and DR data are required to accurately correct for unobserved Kr ions in photoionized nebulae, and hence to determine elemental Kr abundances.  PI cross sections have been determined for ground configuration states of Kr$^0$--Kr$^{5+}$ up to 100~Rydbergs.  Our Kr$^{+}$ PI calculations were significantly improved through comparison with experimental measurements.  RR and DR rate coefficients were respectively determined from the direct and resonant PI cross sections at temperatures (10$^1-10^7$)$z^2$~K, where $z$ is the charge.  We account for $\Delta n=0$ DR core excitations, and find that DR is the dominant recombination mechanism for all but Kr$^+$ at photoionized plasma temperatures.  Internal uncertainties are estimated by comparing results computed with three different configuration-interaction expansions for each ion, and by testing the sensitivity to variations in the orbital radial scaling parameters.  The PI cross sections are generally uncertain by 30--50\% near the ground state thresholds.  Near 10$^4$~K, the RR rate coefficients are typically uncertain by less than 10\%, while those of DR exhibit uncertainties of factors of 2 to 3, due to the unknown energies of near-threshold autoionizing resonances.  With the charge transfer rate coefficients presented in the third paper of this series, these data enable robust Kr abundance determinations in photoionized nebulae for the first time, providing a new tool for studying heavy element enrichments in PNe and for investigating the chemical evolution of trans-iron elements.}


\maketitle

\titlerunning{Atomic Data for Neutron-Capture Elements II}
\authorrunning{N.\ C.\ Sterling}

\section{Introduction} \label{intro}

Recent measurements of neutron(\emph{n})-capture element (atomic number $Z>30$) emission lines in ionized astrophysical nebulae have spurred laboratory astrophysics investigations of these species.  Trans-iron elements were first detected in planetary nebulae (PNe) 35 years ago \citep{treffers76}, but it was not until nearly two decades later that their emission lines were identified \citep{pb94}.  

The detection of these elements marked a new era in the field of PNe, since the progenitor stars of these objects can produce trans-iron elements via slow \emph{n}-capture nucleosynthesis (the \emph{s}-process) during the asymptotic giant branch (AGB) phase of evolution.  Abundance determinations of PNe complement spectroscopic investigations of AGB stars, enabling previously unexplored details of \emph{s}-process nucleosynthesis to be revealed \citep[e.g., see][]{sterling08, sterling11b}.  Indeed, many of the \emph{n}-capture elements that are accessible in nebular spectra, such as Se, Kr, and Xe, cannot be detected in cool giants such as AGB stars (and in fact had not previously been studied in any of their sites of origin).

The pioneering work of \citet{pb94} inspired other groups \citep{dinerstein01a, dinerstein01b, sharpee07} to identify previously unrecognized nebular emission lines as transitions of trans-iron elements.  In spite of the low cosmic abundances of \emph{n}-capture elements \citep{asplund09}, deep ultraviolet \citep{sterling02, sterling03, williams08}, optical \citep[e.g.,][]{hyung01, sharpee03, liu04, zhang05, sharpee07, sterling09, otsuka10b, otsuka11} and near-infrared spectroscopy \citep{likkel06, sterling07, sterling08} of several Galactic and Local Group PNe revealed spectral features of these species.  Abundances of these elements have now been estimated in approximately 100 PNe, allowing for an analysis of \emph{s}-process elemental enrichments in a statistically meaningful sample of objects.

These detections are not limited to PNe.  Indeed, \emph{n}-capture element emission lines have been detected in a variety of other environments, including H~II regions \citep{aspin94, lumsden96,luhman98, baldwin00, puxley00, okumura01, blum08, roman-lopes09} and starburst galaxies \citep{vanzi08}.  These detections demonstrate the promise of nebular spectroscopy as a tool to study \emph{n}-capture nucleosynthesis in low-mass stars, the chemical evolution of trans-iron elements in the Universe, and the heavy-element nucleosynthetic histories of other galaxies.

Kr is one of the most readily detected \emph{n}-capture elements in ionized nebulae, along with Se and Xe, with relatively strong transitions in both the optical \citep{pb94} and near-infrared \citep{dinerstein01a}.  It has been detected in dozens of Galactic PNe and H~II regions, as well as a handful of extragalactic nebulae \citep[e.g.,][]{wood07, otsuka11}.

Beyond the relative ease of detection, Kr is a sensitive probe of \emph{s}-process nucleosynthesis.  It can be highly enriched by the \emph{s}-process, due in part to the $^{86}$Kr isotope, which has a magic number of neutrons and acts as a bottleneck in the \emph{s}-process path.  Isotopes with magic numbers of neutrons have very small \emph{n}-capture cross sections and cause the peaks seen in the element-by-element \emph{s}-process enrichment distribution near $Z=40$, 56, and 82 \citep[e.g.,][]{busso99, burris00, sneden08}.  Typically, \emph{s}-process nucleosynthesis enriches Kr by a significantly larger factor than Se \citep{sterling08, karakas09}, which does not have a stable isotope with a magic number of neutrons.  In addition, because it is a noble gas, Kr is not depleted into dust grains, and hence its gas-phase abundance is representative of the total elemental abundance.

However, nebular spectroscopy requires a foundation of atomic data to determine both ionic abundances and ionization balance solutions to correct for unobserved ions.  Transition probabilities and effective collision strengths have been determined for many detected \emph{n}-capture element ions \citep{biemont86a, biemont86b, biemont87, biemont88, biemont95, schoning97, sb98}, enabling their ionic abundances to be derived.  However, one of the primary hurdles to nebular studies of trans-iron elements is that the ionization equilibria of these species cannot be accurately solved due to the lack of photoionization and recombination data for nearly all \emph{n}-capture element ions.

Motivated by this clear need for atomic data, we have embarked on a program using both theoretical and experimental methods to determine photoionization (PI) cross sections and rate coefficients for radiative recombination (RR), dielectronic recombination (DR), and charge transfer (CT).  Collisional ionization and high-temperature DR are of negligible importance for ionization balance solutions in photoionized nebulae such as PNe and H~II regions.

This paper is the second in a series presenting atomic data calculations for \emph{n}-capture element ions.  In the first \citep{sterling11b}, multi-configuration Breit-Pauli (MCBP) distorted-wave PI cross sections and RR and DR rate coefficients were given for the first six ions of Se.  We furnish CT rate coefficients for the first five ions of Ge, Se, Br, Kr, Rb, and Xe in the third paper \citep{sterling11c}, computed in the Demkov and Landau-Zener approximations.  In this study, we present MCBP distorted-wave PI cross sections and RR and DR data for the first six Kr ions, calculated with the atomic structure code AUTOSTRUCTURE \citep{badnell86, badnell97, badnell11}.  We do not investigate more highly-charged Kr species, since PN central stars and massive young stars (the ionizing sources of H~II regions) are not sufficiently hot to significantly photoionize Kr more than six times \citep[Kr$^{6+}$ has an ionization potential of 111~eV, and PN central stars generally have effective temperatures less than 200\,000~K;][]{napiwot99, phillips03, villaver07}.

Photoionization and recombination studies of low-charge Kr ions are few and far between, particularly in terms of theoretical calculations.  On the other hand, neutral Kr has been the subject of numerous experimental \citep[e.g.,][and references therein]{henke67, lang75, marr76, west76, samson89, samson91, sorokin00} and theoretical \citep[e.g.,][]{amusia71, miller77, parpia84, tulkki92} PI studies.  However, neutral Kr is a trace species in PNe and H~II regions, and therefore its PI properties are not expected to play a significant role in Kr ionization balance solutions.

In terms of recombination, little is known about the ions of interest in our work.  \citet{mattioli06} compiled RR and DR rate coefficients for all Kr ions, but the data for low-charge ions are rough estimates.  For these species, DR rate coefficients were computed using the Burgess formula \citep{burgess65}, as modified by  \citet{merts76}, which is appropriate for collisionally-ionized plasmas (indeed, the work of Mattioli et al.\ was inspired by the utility of Kr as a coolant for magnetically-confined fusion plasmas), but is known to badly underestimate the rate coefficient at the low temperatures of photoionized plasmas.  Their RR rate coefficients for low-charge Kr ions are approximate, with RR into the valence and (hydrogenic) excited shells treated separately.

Experimental measurements of absolute PI cross sections provide the most useful comparisons to our calculations.  Concurrent to our theoretical investigations, we have experimentally measured PI cross sections for Se and Xe ions \citep{esteves09, esteves10, sterling11a, esteves11a, esteves11b} at the Advanced Light Source synchrotron radiation facility, for the purpose of constraining our calculations.  PI investigations were previously carried out for Kr$^{3+}$ to Kr$^{5+}$ \citep{lu06a, lu06b, lu_thesis}, and more recently for Kr$^+$ \citep{bizau11}.  We compare our computed PI cross sections to these measurements to benchmark and gauge the accuracy of our direct PI cross sections.

Because no comprehensive studies of the PI and recombination properties of low-charge Kr ions in the photoionized regime existed prior to our study, we provide estimates for the internal uncertainties of our computed atomic data.  These uncertainties can affect ionization equilibrium calculations, and hence elemental abundance determinations.  Beyond comparisons to experimental PI cross-section measurements, we computed data using three different configuration-interaction (CI) expansions for each Kr ion to gauge the sensitivity of the results to CI effects.  Moreover, we tested the effects of other parameters and assumptions in our calculations (e.g., whether the radial wavefunctions were orthogonalized, the radial scaling parameters for continuum orbitals, etc.) on the resulting data.

In the remainder of this paper, we present the computed PI cross sections and RR and DR rate coefficients for low-charge Kr ions.  In Sect.~\ref{calcs}, we briefly describe the calculations and methodology for each of these processes, and detail the calculated electronic structure for each ion.  The resulting data are presented in Sect.~\ref{results}, along with estimated uncertainties and comparisons to previous studies.  We summarize our investigation in Sect.~\ref{summary}.

\section{Calculations and methodology}\label{calcs}

The electronic structure, distorted-wave PI cross sections, and RR and DR rate coefficients have been computed for the first six ions of Kr using the atomic structure code AUTOSTRUCTURE \citep[v23.25;][]{badnell86, badnell97, badnell11}.  We refer the reader to \citet{badnell03} and \citet{badnell06b} for a detailed account of the theoretical background for these calculations..

\subsection{Electronic structure}

We constructed three separate CI expansions (which we name ``small,'' ``medium,'' and ``large'') for each ion, in order to test the sensitivity of our results to the adopted CI expansion.  Based on comparisons to level energies and ionization potentials in NIST \citep{nist}, the medium configuration sets provide the best compromise between accuracy and computational efficiency, and the results in Sect.~\ref{results} were calculated with the medium CI expansions.  In Table~\ref{ciexp},\footnote{Tables~1-4 are available in the online version of this article.} the configurations and numbers of levels for the three CI expansions of each ion are given.

The electronic structure of each ion was computed with Thomas-Fermi-Dirac-Amaldi potentials, in intermediate coupling with $\kappa$-averaged relativistic wavefunctions \citep{cowan76}.  The radial wavefunctions were Schmidt orthogonalized, but calculations for each process were also conducted without orthogonalization (see Sect.~\ref{results}).  We determined orbital radial scaling parameters (Table~\ref{lambdas}) by optimizing on the average of LS term energies in order to best reproduce experimental energies and ionization potentials tabulated by NIST.   We generally found best agreement with experimental values by optimizing the scaling parameter for each orbital on terms in the lowest-energy configuration containing electrons in the that orbital.  In a few cases, other optimization strategies were required to improve the accuracies of ground configuration level energies and the ionization potential (e.g., see the discussion of the Kr$^+$ structure in Sect.~\ref{pi_exp}).  The computed energies for the medium and large configuration sets typically agree with experimental determinations to within 3--4\% (see Table~\ref{ecomp}), while the small CI expansion energies are in slightly worse agreement.  The energies for some levels are more discrepant, especially $^1$D$_2$ and $^2$D$_{3/2, 5/2}$ in the ground configurations of Kr$^{2+}$--Kr$^{4+}$ (the disagreement is as high as 22\% for the medium CI expansion of Kr$^{3+}$ 4s$^2$\,4p$^4$~$^2$D$_{3/2}$).  

Calculated ionization potentials (Table~\ref{ecomp}) agree with NIST values to within 2\% for the medium and large configuration sets, with the exceptions of Kr$^{2+}$ and Kr$^{3+}$.  The ionization potential of Kr$^{2+}$ is strongly sensitive to the radial scaling parameter of the $4p$ orbital.  Decreasing the $4p$ scaling parameter leads to an improved ionization potential, but the accuracy of the ground term energies rapidly worsens.  If the $4p$ scaling parameter optimization is conducted so as to minimize its value, the computed ionization potential would improve from an accuracy of 7.1\% to 6.2\%, while the errors in the $^3$P metastable state energies would more than triple.  We therefore chose to optimize the $4p$ scaling parameter to best reproduce the experimental energy level structure, at the expense of a slightly less accurate ionization potential.  This choice does not significantly affect our results -- the direct Kr$^{2+}$ PI cross section computed with the two values of the $4p$ scaling parameter differs by less than 2\% near the ground state threshold.  The Kr$^{3+}$ ionization potential is quite sensitive to the $4s$ and $4p$ scaling parameters, improving when those values are decreased.  However, in this case it was not possible to optimize the $4s$ and $4p$ scaling parameters so as to reduce their values below those listed in Table~\ref{lambdas}, and hence we were unable to further improve the accuracy of the ionization potential.




In Table~\ref{acomp}, we compare the Einstein $A$-coefficients we computed with previously determined values.  For forbidden transitions in the ground configuration, our $A$-coefficients agree to within 50\% or better with previous results, with the exception of transitions involving levels whose energy we were not able to compute to within 3--4\% accuracy (e.g., $^1$D$_2$ for Kr$^{2+}$ and Kr$^{4+}$, and $^2$D$_{3/2, 5/2}$ for Kr$^{3+}$).  For permitted transitions, the disagreement is generally larger (especially for neutral and singly-ionized Kr).

\subsection{Photoionization, radiative recombination, and dielectronic recombination}

In low-density nebulae such as PNe and H~II regions, nearly all atomic ions reside in ground configuration states.  Therefore, we computed MCBP distorted-wave PI cross sections for each level of the Kr$^0$ to Kr$^{5+}$ ground configurations from the ionization threshold up to 100~Rydbergs.  Likewise, RR and DR rate coefficients were computed for states within the ground configuration of the target ions (i.e., before recombination) Kr$^+$ to Kr$^{6+}$ over the temperature range (10$^1-10^7$)$z^2$~K.  The RR and DR rate coefficients were determined from the direct and resonant portions of the PI cross sections (respectively), using detailed balance.  For DR, only $\Delta n=0$ core excitations were considered, since these dominate $\Delta n\geq 1$ core excitations at photoionized plasma temperatures.  However, for DR forming Kr$^0$ and Kr$^+$, whose lowest energy 4$s^2$\,4$p^k$\,5$s$ and (in the case of Kr$^0$) 4$s^2$\,4$p^k$\,5$p$ levels ($k=4$ and 5 for Kr$^+$ and Kr$^0$, respectively) lie below those of 4$s^2$\,4$p^k$\,4$d$, we allowed $\Delta n=1$ excitations into the $5s$ and $5p$ orbitals.  We ignored the small contributions of core excitations into $5d$, $5f$, and $5g$ orbitals to keep the size of the calculations tractable.

The CI expansions listed in Table~\ref{ciexp} were used in these calculations, with the $(N+1)$-electron ion scaling parameters from Table~\ref{lambdas}.  Our tests using target ($N$-electron) scaling parameters led to substantial disagreement between our calculated PI cross sections and experimental measurements in all cases.  The scaling parameters of $s$, $p$, and $d$ continuum orbitals are assumed to be the same as those of the highest principal quantum number bound orbitals with the same orbital angular momentum, and equal to unity otherwise.  For RR, the $(N+1)$-electron ion's CI expansion was augmented to include one-electron additions to the target's ground and mixing configurations.  In the case of DR, all relevant core excitations were added to the target configuration set, and the $(N+1)$-electron ion's CI expansion was expanded to include one-electron additions to target states to allow Rydberg electrons to radiate into the core.

AUTOSTRUCTURE automatically switches from length to velocity to acceleration gauge as the photon energy increases.  However, because we employed $\kappa$-averaged relativistic radial wavefunctions, acceleration gauge could not be used and it was necessary to force velocity gauge at high energies.  In Sects.~\ref{pi_error} and \ref{rr_error}, we show that this does not affect our calculations at the energies and temperatures of interest.  

RR and DR were treated separately using the independent processes approximation \citep{pindzola92}.  In our RR calculations, we used a maximum principal quantum number of 1000, and a maximum orbital angular momentum of $l=4$ for the first two ions and $l=3$ for more highly charged species (hydrogenic radial integrals from the recurrence relations of \citet{burgess64} were used for higher values of $l$).  For DR, we used a maximum $n=1000$ and $l=200$, with $l\geq 9$ treated hydrogenically except for the Kr$^{5+}$ and Kr$^{6+}$ targets, for which $l\geq 10$ and 11 (respectively) were computed hydrogenically in order to allow the calculation to converge.  To improve the accuracy of our DR calculations, we utilized experimental target energies whenever available, with theoretical level splittings for terms with incomplete energy information in NIST \citep{badnell06a}.

\section{Results}\label{results}

\subsection{Photoionization cross sections}

\begin{figure}
  \resizebox{\hsize}{!}{\includegraphics{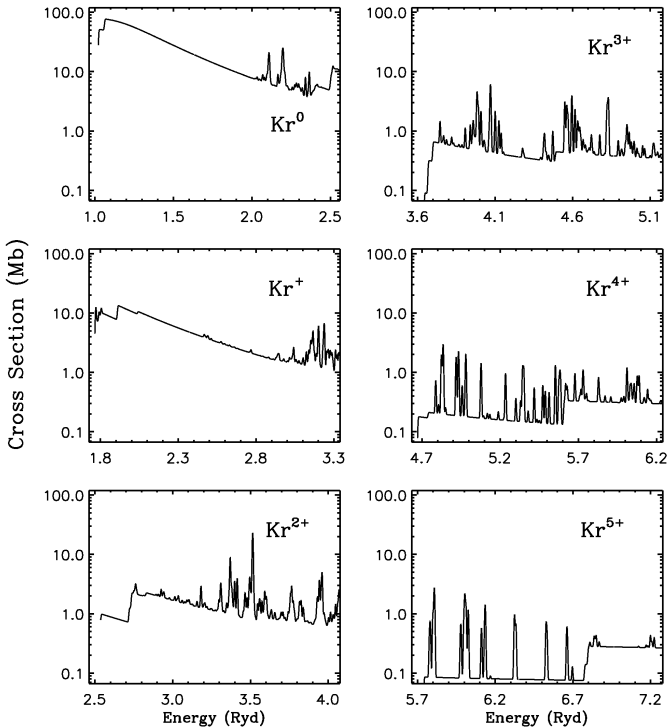}}
  \caption{Ground state PI cross sections for Kr$^0$--Kr$^{5+}$ near their ionization thresholds.} \label{allpi}
\end{figure}

In Figure~\ref{allpi}, we illustrate the ground state Kr$^0$ to Kr$^{5+}$ PI cross sections near their ionization thresholds.  The AUTOSTRUCTURE output was processed with the codes ADASPI (for direct PI) and ADASPE (for photoexcitation-autoionization), which produce the ADAS \textit{adf39} and \textit{adf38} output files, respectively \citep{summers05}.  For ease of presentation, we used the post-processing code \textit{xpeppi} to add the direct and resonant portions of the PI cross sections.

The PI cross sections are available at the CDS, with two data files for each ion.  Files labeled ``kr$q$+\_XPITOT.dat'' (where $q$ is the charge of the $(N+1)$-electron ion) tabulate the direct-plus-resonant PI cross sections in five columns.  Photon energies relative to the ground and initial state are given in the first two columns (respectively), while the next two columns present the energies of the photo-electron relative to the ground and initial states.  All energies are in Rydbergs.  The cross section in Mb is given in the final column.  For ions with multiple ground configuration levels, the cross sections for different levels are delimited by comment lines (indicated by a ``\#'' in the first column) followed by three numbers.  The first is the energy of the initial level relative to the ground state (or the total energy, in the case of the ground state itself), the second is the total energy of the final target state, and the last gives the energy order of the initial level (1 for the ground state, 2 for the first excited state, etc.).

Direct PI cross sections, without photoexcitation-autoionization resonances, are presented in files labeled ``kr$q$+\_XDPITOT.dat.''  These files contain only two columns, the first of which is the photon energy in Rydbergs, and the second the cross section in Mb.  Cross sections for different states are delimited as in the kr$q$+\_XPITOT.dat files.

\subsubsection{Estimation of uncertainties}\label{pi_error}


We estimate uncertainties by testing the sensitivity of our results to the orbital radial scaling parameters, the adopted CI expansion,  and to various assumptions in our calculations.  We also compare our results to experimental absolute PI cross-section measurements (Sect.~\ref{pi_exp}).

\begin{figure}
  \resizebox{\hsize}{!}{\includegraphics{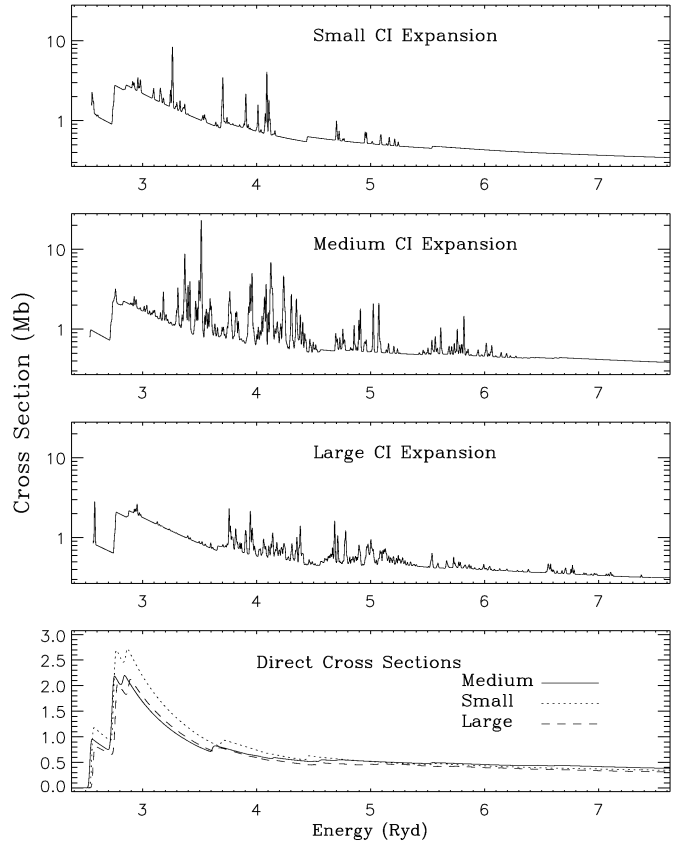}}
  \caption{Comparison of Kr$^{2+}$ ground state PI cross sections computed for each of the three CI expansions (as indicated in the top three panels).  The direct PI cross sections are compared in the bottom panel.} \label{kr2comp}
\end{figure}

\begin{figure}
  \resizebox{\hsize}{!}{\includegraphics{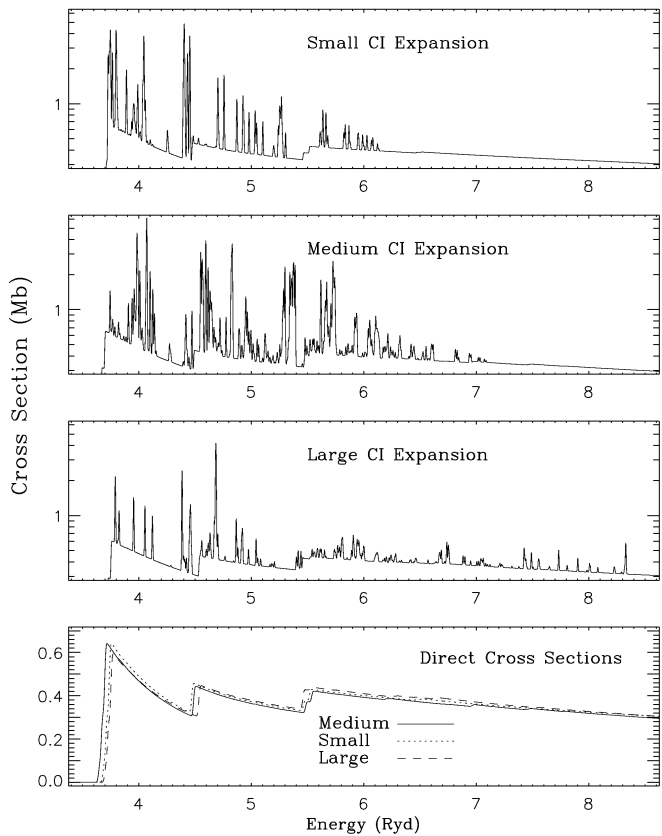}}
  \caption{Comparison of Kr$^{3+}$ ground state PI cross sections computed for each of the three CI expansions (as indicated in the top three panels).  The direct PI cross sections are compared in the bottom panel.} \label{kr3comp}
\end{figure}

The PI cross sections are most sensitive to the adopted CI expansion.  In Figs.~\ref{kr2comp} and \ref{kr3comp}, we show the Kr$^{2+}$ and Kr$^{3+}$ cross sections computed with each of the three configuration sets (top three panels), while the bottom panel compares the direct cross sections from each calculation.  Most notably, the resonance structure, both in terms of strengths and positions, varies considerably among the different CI expansions.  This illustrates the challenge of modeling photoexcitation-autoionization processes in the distorted-wave approximation.  It should be noted that any theoretical treatment leads to uncertainties in PI resonance structure, so much so that codes used to numerically simulate astrophysical nebulae often ignore these resonances \citep{ferland98} or use cross sections convolved with a broad Gaussian to smooth out uncertainties in resonance positions \citep{kb01}.

We therefore focus on estimating uncertainties in the direct PI cross sections, particularly near the ground state ionization threshold.  The Kr$^{2+}$ cross section computed with the small configuration set is larger than those from the other CI expansions by approximately 25\% near the ground state threshold (Fig.~\ref{kr2comp}), while the Kr$^{3+}$ direct cross sections agree to within 5\% (Fig.~\ref{kr3comp}).  Likewise, the direct cross section computed with the medium Kr$^{4+}$ and Kr$^{5+}$ CI expansions agree with those from the small and large configuration sets to within 5--10\%.  The disagreement is largest (up to $\sim$40\%) for neutral and singly ionized Kr.

Other tests revealed smaller uncertainties in the direct PI cross sections.  For example, setting the scaling parameters to unity for $s$, $p$, and $d$ continuum orbitals (rather than the value of the highest principal quantum number bound orbital with the same orbital angular momentum) led to differences of 10\% near the ground state threshold of neutral Kr, and less than 5\% for other Kr ions.  If the PI cross sections are computed without Schmidt orthogonalizing the radial orbitals, the direct cross sections differ by 5\% or less in all cases.  To test whether our inability to switch to acceleration gauge at high energies (due to the use of $\kappa$-averaged relativistic orbitals) leads to inaccuracies, for each ion we ran a PI calculation solely in the length gauge, and another solely in the velocity gauge.  The cross sections showed negligible difference ($\ll 1$\%), indicating that our PI calculations are not affected by forcing velocity gauge up to 100~Ryd.

Based on these tests and comparisons to experimental measurements, we estimate the uncertainties in our direct PI cross sections to be $\sim$30--50\% for most Kr ions.  However, comparison with experimental measurements (Sect.~\ref{pi_exp}) indicate that the uncertainties in the Kr$^0$ and Kr$^{3+}$ PI cross sections may be as large as a factor of two.

\subsubsection{Comparison to experiment} \label{pi_exp}

It is critical to benchmark atomic data calculations against experimental measurements whenever possible.  Of the six Kr species considered, experimental absolute PI cross sections have been measured for all but Kr$^{2+}$.

The measurements were conducted at a variety of synchrotron radiation facilities, and in the case of ionized species utilized the merged beams method.  Kr$^{3+}$ through Kr$^{5+}$ PI cross sections were measured at the Advanced Light Source synchrotron radiation facility at the Lawrence Berkeley National Laboratory in California \citep{lu06a, lu06b, lu_thesis}, while that of Kr$^+$ was recently measured at the ASTRID storage ring at the University of Aarhus in Denmark \citep{bizau11}.  These measurements were complicated by the population of metastable states in the primary ion beams, and hence the measured cross sections are a linear combination of those from the ground and metastable states.  However, \citet{bizau11} used an ion trap to significantly limit contributions from the $^2$P$_{1/2}$ metastable state, and measured the Kr$^+$ ground state PI cross section over a limited energy range.  In the case of neutral Kr, we compare our results to the PI cross sections recommended by \citet{richter03}, which were determined from measurements of the PI and electron-impact ionization cross section ratio and normalized to the double ion chamber PI measurements of \citet{samson89} and \citet{samson91}.

\begin{figure}
  \resizebox{\hsize}{!}{\includegraphics{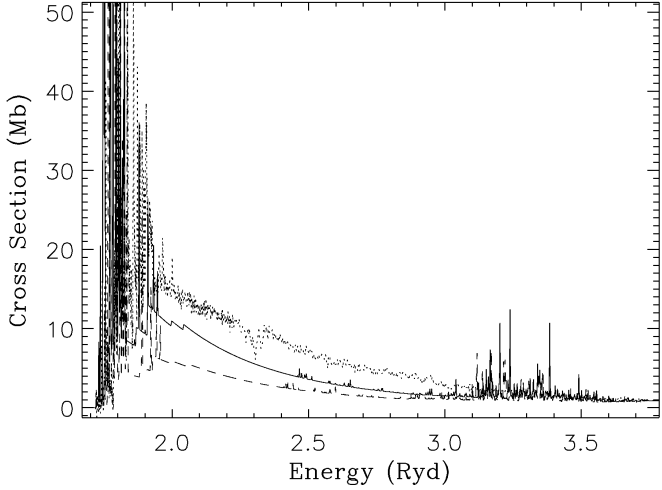}}
  \caption{Comparison of calculated Kr$^+$ PI cross section to the experimental measurements (dotted line) of \citet{bizau11}.  The dashed line shows our original result before comparison to experiment, while the solid line shows the theoretical cross section after optimization to better reproduce the \citet{bizau11} results.  In each case, the theoretical curves are statistically-weighted linear combinations of the cross sections from each of the two states in the ground configuration.  The calculated cross sections were convolved with a Gaussian of 30~meV to approximate the average resolution of the \citet{bizau11} measurements.}\label{kr1exp}
\end{figure}

We substantially improved our calculated Kr$^+$ PI cross section through comparison with the experimental data of \citet{bizau11}.  In Fig.~\ref{kr1exp}, the experimental measurements are shown as a dotted line, and the dashed line represents our original calculation -- statistically weighted and convolved with a Gaussian of 30~meV (the average experimental photon energy resolution near the ground state threshold).  The calculated cross section was originally a factor of 2.5 smaller than the experimental one.  We added and removed several configurations from the target and $(N+1)$-electron ion CI expansions, none of which improved agreement with experiment.  

Tests showed the PI cross section to be sensitive to the $4p$, $4d$, $5p$, and continuum $d$ orbital radial scaling parameters, generally increasing (improving) with smaller values for those parameters.  We therefore re-optimized the Kr$^+$ $np$ and $nd$ scaling parameters so as to minimize their values without significantly harming the agreement with the experimental level energies and ionization potential.  In our original structure calculation, the $4p$ scaling parameter was optimized on terms in the ground, $4s$\,$4p^6$, $4s^2$\,$4p^4$\,$4d$, and $4s^2$\,$4p^4$\,$5s$ configurations and the $5p$ scaling parameter was optimized on $4s^2$\,$4p^3$\,$5p^2$ terms (this configuration was later removed from the CI expansion) in order to reproduce the energy of the first excited state ($^2$P$_{1/2}$).  The $4d$ scaling parameter was optimized on $4s^2$\,$4p^3$\,$4d^2$ terms, to best reproduce the ionization potential.  In our new optimization, we instead optimized the $4p$ radial scaling parameter only on the ground configuration term, and those for $4d$ and $5p$ on terms in the lowest-energy configurations containing an electron in those orbitals ($4s^2$\,$4p^4$\,$4d$ and $4s^2$\,$4p^4$\,$5p$, respectively).  This new optimization reduced the accuracy of the first excited state energy from 0.8\% to 11\% (higher-energy levels are not appreciably affected), and that of the ionization potential from 0.5\% to 1.5\%.  However, it produced a significantly larger direct PI cross section (solid line in Fig.~\ref{kr1exp}), within 40\% of the experimental data.  This level of improvement demonstrates how useful absolute experimental PI cross sections can be to theoretical determinations.

\begin{figure}
  \resizebox{\hsize}{!}{\includegraphics{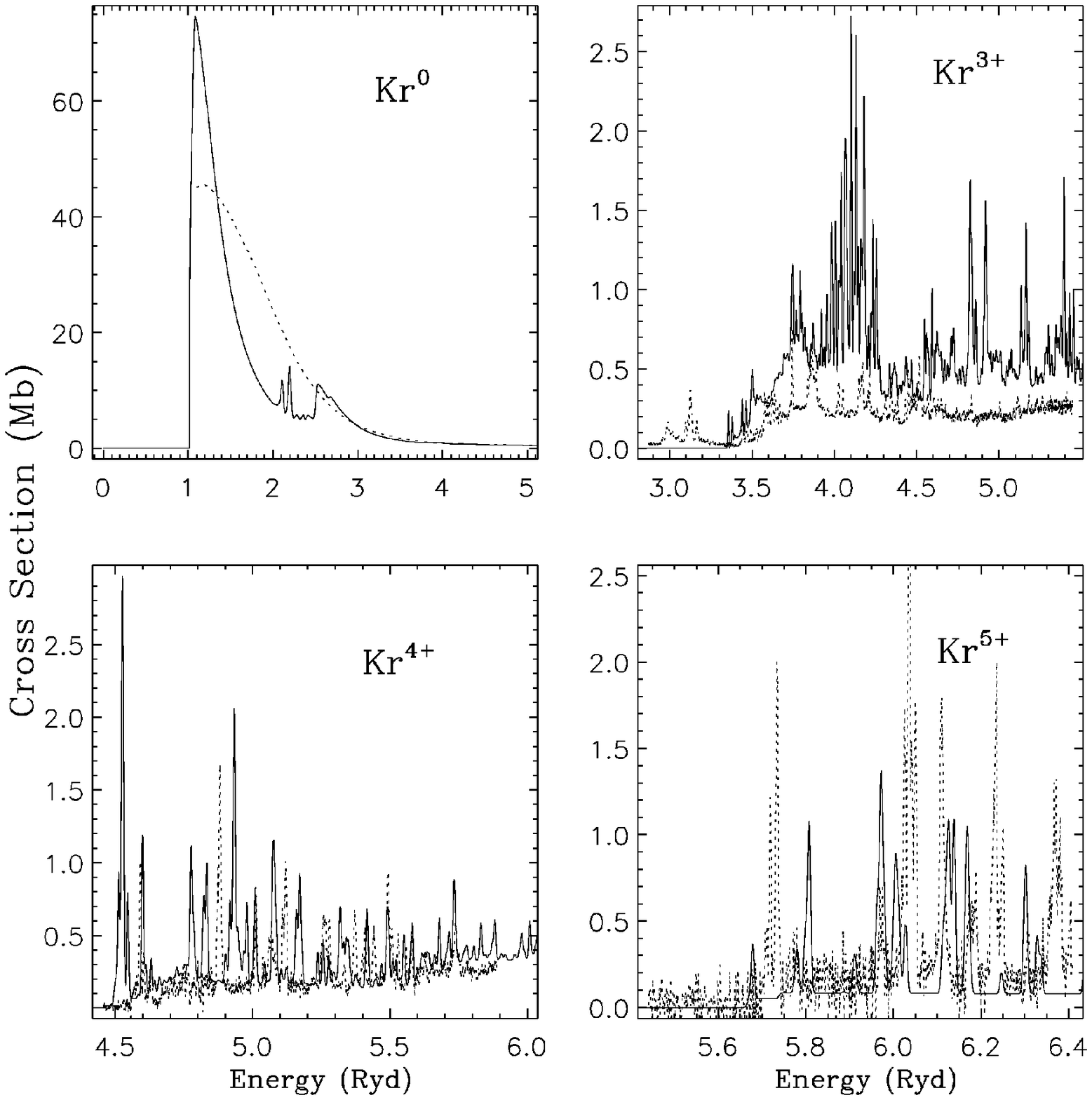}}
  \caption{Comparison of calculated PI cross sections (solid lines) with experimental data (dotted lines) for Kr$^0$ \citep{richter03}, Kr$^{3+}$ \citep{lu06b}, Kr$^{4+}$ \citep{lu_thesis}, and Kr$^{5+}$ \citep{lu06a}.  The theoretical curves have been convolved with a Gaussian of 1~eV for Kr$^0$ and 100~meV for the other species, to match the experimental resolution.  The theoretical data are linear combinations of PI cross sections from each state in the ground configuration, weighted according to statistical weights in the case of Kr$^{3+}$, and by the factors ($^3$P$_0$, $^3$P$_1$, $^3$P$_2$, $^1$D$_2$, $^1$S$_0$)~=~(0.32, 0.15, 0.2, 0.3, 0.03) for Kr$^{4+}$ and ($^2$P$_{1/2}$, $^2$P$_{3/2}$)~=~(0.4, 0.6) for Kr$^{5+}$.} \label{krexp_panel}
\end{figure}

We compare our PI cross sections to experimental measurements for other Kr ions in Fig.~\ref{krexp_panel}.  The depicted theoretical data are linear combinations of cross sections for states in the ground configurations that were visually determined to best reproduce the experimental direct PI cross section and, when possible, families of resonances.  The cross sections were weighted by the factors ($^3$P$_0$, $^3$P$_1$, $^3$P$_2$, $^1$D$_2$, $^1$S$_0$)~=~(0.32, 0.15, 0.20, 0.3, 0.03) for Kr$^{4+}$, ($^2$P$_{1/2}$, $^2$P$_{3/2}$)~=~(0.4, 0.6) for Kr$^{5+}$, and by statistical weights in the case of Kr$^{3+}$.  The calculated cross sections were convolved with a Gaussian with FWHM equal to the experimental resolution ($\sim$1~eV for Kr$^0$, and 0.1~eV for the other three ions).

In the cases of Kr$^{4+}$ and Kr$^{5+}$, the direct cross sections generally agree within the experimental noise.  The theoretical cross section is slightly too large for Kr$^{4+}$ and perhaps slightly small for Kr$^{5+}$, but over the full range of energies they agree to within the experimental uncertainties of $\sim$25\%.


The agreement is markedly worse for Kr$^{3+}$, for which the theoretical cross section is larger than that measured experimentally  by approximately a factor of two.  Changing the metastable fractions does not significantly improve the agreement with the experimental measurements, since the calculated direct PI cross sections of the ground and metastable states have comparable magnitudes and hence any linear combination of their cross sections will result in poor agreement with the experimental data.  As seen in Fig.~\ref{kr3comp}, the direct Kr$^{3+}$ PI cross section is not strongly sensitive to the CI expansion.  We verified this by adding and removing several configurations in the target and $(N+1)$-electron ion medium configuration sets, none of which significantly affected the direct PI cross section.  Near the ground state threshold, the direct cross section is sensitive to the values of the $4d$ and continuum $s$ and $d$ scaling parameters, decreasing (coming into better agreement with experiment) as the values of those scaling parameters increase.  However, our re-optimization designed to increase the $4d$ (and hence continuum $d$) scaling parameter -- by optimizing on $4s^2$\,$4p$\,$4d^2$ instead of $4s^2$\,$4p^2$\,$4d$ terms -- did not effect changes in the direct PI cross section of more than 10--15\%.  Moreoever, while we were able to reduce the direct PI cross section near threshold, the cross section above 4.5~Ryd increased (worsening agreement with experiment) by 10\%.  Therefore, we did not alter our calculation, and concede that the computed Kr$^{3+}$ PI cross section is uncertain by approximately a factor of two.

Likewise, for Kr$^0$ the computed cross section is too large at threshold by $\sim$70\% compared to experiment.  The cross section falls below experiment near 1.35~Ryd, and the disagreement increases to a factor of two until $\sim$2.5~Ryd, where a threshold corresponding to Kr$^+$ 4$s^2$\,4$p^4$\,5$p$ is seen in the theoretical cross section.  The PI cross section at threshold is very sensitive to the value of the $4p$ and $5p$ scaling parameters, but altering their values in the sense required to improve the direct cross section badly damages energies of excited levels and the ionization threshold.  Like Kr$^{3+}$, the calculated PI cross section is uncertain by a factor of two, but we do not expect this to harm ionization balance solutions significantly since neutral Kr is a trace species in ionized astrophysical nebulae.



\subsection{Radiative recombination rate coefficients}

Total and partial final-state resolved RR rate coefficients were derived from the direct portion of the PI cross sections using detailed balance.  We used the ADASRR code to process the AUTOSTRUCTURE output and produce ADAS \textit{adf48} files (available at the CDS), which are similar in structure to \textit{adf09} files for DR \citep{summers05}.  At the end of the \textit{adf48} files, total RR rate coefficients for all ground configuration states of the target ion are printed.  Information preceding those rates pertain to partial RR rate coefficients, which are fully $J$-resolved for $n\leq 8$ levels, bundled-$nl$ for $n\leq 10$ states, and bundled$-n$ for all higher levels \citep[see][]{badnell06a}.

The rate coefficients were fit with the analytical formula \citep{verner96a, badnell06b}
\begin{equation}
\alpha_{\mathrm{RR}}(T) = A \times \left[ \sqrt{T/T_0}(1 + \sqrt{T/T_0})^{1-B_0}(1 + \sqrt{T/T_1})^{1+B_0} \right]^{-1}, \label{rr1}
\end{equation}
where 
\begin{equation}
B_0 = B + C\mathrm{exp}(-T_2/T). \label{rr2}
\end{equation}
The fit coefficients $T$ and $T_{0,1,2}$ are in temperature units, $A$ and $\alpha_{\mathrm{RR}}(T)$ are in cm$^3$\,s$^{-1}$, and $B$ and $C$ are dimensionless.

As was the case for Se ions \citep{sterling11b}, we found that Equations~\ref{rr1} and \ref{rr2} do not always provide satisfactory fits to the Kr RR rate coefficients over the full temperature range (10$^1-10^7$)$z^2$~K.  The reason is that the slopes of the rate coefficients can increase significantly at high temperatures.  These ``bumps'' are caused by electron capture into low-energy levels, which are not as effectively screened from the nucleus as higher levels.  Equations~\ref{rr1} and \ref{rr2} can be used to fit the low-temperature behavior of the RR rate coefficients, but a different analytical form is often necessary to fit the high-temperature behavior.  For these high-temperature fits, we have chosen to utilize the formula typically used to fit DR rate coefficients \citep[e.g., ][]{zatsarinny03}:
\begin{equation}
\alpha_{\mathrm{RR}}(T) = \frac{1}{T^{3/2}} \sum_{i=1}^{n} c_i\mathrm{exp}(-E_i/T), \label{dreq}
\end{equation}
where $T$ and $E_i$ are in temperature units, $c_i$ are in cm$^3$\,s$^{-1}$\,K$^{3/2}$, and $\alpha_{\mathrm{RR}}(T)$ is in cm$^3$\,s$^{-1}$.  The value for $n$ in Equation~\ref{dreq} ranges from 5 to 7, depending on the target state.

The fit coefficients for the RR rate coefficients are presented in Table~\ref{rrlowtfits} for the low-temperature fits (using Equations~\ref{rr1} and \ref{rr2}) and Table~\ref{rrhitfits} for the high-temperature fits (using Equation~\ref{dreq}).  Table~\ref{rrlowtfits} also lists the maximum temperature at which the low-temperature fits are valid for each state; if $T_{\rm max}$ is not given, then the low-temperature fits are valid over the entire temperature range (10$^1-10^7$)$z^2$~K.  Unless noted otherwise, the fits are accurate to within 5\% at low temperatures (Table~\ref{rrlowtfits}) and 2\% at high temperatures (Table~\ref{rrhitfits}), and respectively reproduce the correct low-temperature and high-temperature asymptotic behaviors of the rate coefficients.

\addtocounter{table}{4}

\begin{table*}
\centering
\caption{Fit coefficients for radiative recombination rate coefficients at low temperature (see Equations~\ref{rr1} and \ref{rr2}).  }\label{rrlowtfits} 
\begin{tabular}{lcccccccc}
\hline \hline
Target & & $A$ & $B$ & $T_0$ & $T_1$ & $C$ & $T_2$ & $T_{\rm max}$ \\
Ion\tablefootmark{a} & Level & (cm$^3$\,s$^{-1}$) &  & (K) & (K) &  & (K) & (K) \\
\hline
Kr$^+$ & 1\tablefootmark{b} & 1.325(-9) & 0.7589 & 2.544(-2) & 1.404(6) & 0.3345 & 2.594(6) & ... \\
 & 2\tablefootmark{c} & 4.239(-11) & 0.5384 & 2.527 & 4.406(143) & 0.2391 & 6.021(1) & 1.0(3) \\
Kr$^{2+}$ & 1 & 6.988(-11) & 0.7335 & 5.226(1) & 5.218(5) & 0.7838 & 2.467(6) & 2.0(5) \\
 & 2 & 1.561(-11) & 1.1179 & 1.574(2) & 1.959(5) & 1.0651 & 2.382(6) & 8.0(4) \\
 & 3 & 1.148(-11) & 1.1463 & 2.707(2) & 1.892(5) & 1.1620 & 2.364(6) & 8.0(4) \\
 & 4 & 5.245(-12) & 1.2541 & 9.039(2) & 1.791(5) & 1.4433 & 2.345(6) & 8.0(4) \\
 & 5 & 3.382(-11) & 1.0954 & 1.502(1) & 2.904(5) & 0.7487 & 2.433(6) & 8.0(4) \\
Kr$^{3+}$ & 1 & 9.239(-11) & 0.5956 & 1.311(2) & 9.717(5) & 0.8172 & 2.519(6) & 9.0(4) \\
 & 2 & 3.295(-11) & 1.0865 & 3.924(1) & 3.750(5) & 0.8407 & 2.350(6) & 9.0(4) \\
 & 3 & 3.261(-11) & 1.0860 & 3.881(1) & 3.756(5) & 0.8409 & 2.356(6) & 9.0(4) \\
 & 4 & 3.501(-11) & 1.0727 & 2.409(1) & 4.613(5) & 0.7792 & 2.314(6) & 9.0(4) \\
 & 5 & 8.822(-11) & 1.0328 & 4.088 & 6.267(5) & 0.6287 & 2.527(6) & 9.0(4) \\
Kr$^{4+}$ & 1 & 2.270(-11) & 0.0035 & 4.273(3) & 9.944(7) & 0.7085 & 1.294(6) & ... \\
 & 2\tablefootmark{d} & 4.956(-11) & 0.8138 & 2.308(2) & 8.365(5) & 0.8931 & 2.473(6) & 8.0(4) \\
 & 3\tablefootmark{d} & 4.893(-11) & 0.8718 & 1.507(2) & 8.743(5) & 0.8371 & 2.459(6) & 8.0(4) \\
 & 4\tablefootmark{d} & 3.185(-11) & 0.8859 & 1.474(2) & 1.756(6) & 0.7258 & 2.327(6) & 8.0(4) \\
 & 5\tablefootmark{d} & 6.490(-13) & 0.0000 & 1.344(5) & 1.892(7) & 1.6771 & 9.587(5) & ... \\
Kr$^{5+}$ & 1 & 4.740(-11) & 0.1253 & 2.425(3) & 1.173(8) & 0.5711 & 1.247(6) & ... \\
 & 2\tablefootmark{b} & 3.434(-12) & 0.0000 & 5.369(4) & 5.256(7) & 1.0250 & 1.084(6) & ... \\
Kr$^{6+}$ & 1 & 1.132(-10) & 0.2910 & 1.024(3) & 1.111(8) & 0.4065 & 1.320(6) & ... \\
\hline
\end{tabular}
\tablefoot{$T_{\rm max}$ represents the maximum temperature for which these fits are valid to within the stated accuracies (5\%, unless noted).  For recombination at temperatures exceeding $T_{\rm max}$, the fits from Equation~\ref{dreq} and coefficients listed in Table~\ref{rrhitfits} should be used.  If no $T_{\rm max}$ is given, then the fits from Equations~\ref{rr1} and \ref{rr2} are accurate over the entire temperature range (10$^1-10^7$)$z^2$~K.  The notation $x(y)$ denotes $x\times 10^y$.
\tablefoottext{a}{Note that the ion and level numbers correspond to the target ion (i.e., before recombination).}
\tablefoottext{b}{The fit is accurate to within 6\%}
\tablefoottext{c}{The fit to the Kr$^+$ level 2 RR rate coefficient is inaccurate at the lowest temperatures (by 6\% at 20~K and 17\% at 10~K), and should not be extrapolated to temperatures below 20~K.}
\tablefoottext{d}{For excited states of the Kr$^{4+}$ target, the fits are accurate to within 8\% (level~2), 8.5\% (level~3), 9\% (level~4), and 9.5\% (level~5).}
}
\end{table*}

\begin{table*}
\centering
\caption{Fit coefficients for radiative recombination rate coefficients at high temperatures, valid from $T_{\rm max}$ (see Table~\ref{rrlowtfits}) to $10^7z^2$~K.  See Equation~\ref{dreq}.}\label{rrhitfits} 
\begin{tabular}{lcccccccc}
\hline \hline
Target & &  &  &  &  &  &  &  \\
Ion\tablefootmark{a} & Level & $c_1$ & $c_2$ & $c_3$ & $c_4$ & $c_5$ & $c_6$ & $c_7$ \\
\hline
Kr$^+$ & 2 & 1.412(-7) & 6.899(-7) & 2.071(-6) & 3.400(-6) & 2.266(-5) & 1.824(-4) & 4.196(-4) \\
Kr$^{2+}$ & 1 & 5.457(-6) & 1.743(-5) & 2.608(-4) & 1.151(-3) & 2.277(-3) & ... & ... \\
 & 2 & 2.197(-6) & 9.464(-6) & 1.462(-5) & 2.691(-4) & 1.100(-3) & 2.238(-3) & ... \\
 & 3 & 2.178(-6) & 9.334(-6) & 1.484(-5) & 2.679(-4) & 1.098(-3) & 2.262(-3) & ... \\
 & 4 & 1.952(-6) & 8.700(-6) & 1.405(-5) & 2.540(-4) & 1.043(-3) & 2.112(-3) & ... \\
 & 5 & 1.708(-6) & 7.818(-6) & 1.370(-5) & 2.480(-4) & 1.025(-3) & 2.082(-3) & ... \\
Kr$^{3+}$ & 1 & 1.632(-5) & 2.080(-5) & 1.864(-4) & 1.229(-3) & 3.388(-3) & 5.306(-3) & ... \\
 & 2 & 4.857(-6) & 1.808(-5) & 1.484(-4) & 1.136(-3) & 3.187(-3) & 4.937(-3) & ... \\
 & 3 & 4.766(-6) & 1.771(-5) & 1.449(-4) & 1.107(-3) & 3.069(-3) & 4.796(-3) & ... \\
 & 4 & 9.794(-6) & 1.446(-5) & 1.634(-4) & 1.084(-3) & 3.043(-3) & 4.693(-3) & ... \\
 & 5 & 9.872(-6) & 1.436(-5) & 1.623(-4) & 1.083(-3) & 3.037(-3) & 4.651(-3) & ... \\
Kr$^{4+}$ & 2 & 3.235(-5) & 6.249(-5) & 7.465(-4) & 3.098(-3) & 6.678(-3) & 8.624(-3) & ... \\
 & 3 & 3.007(-5) & 6.022(-5) & 7.165(-4) & 2.946(-3) & 6.454(-3) & 8.442(-3) & ... \\
 & 4 & 2.334(-5) & 5.103(-5) & 6.940(-4) & 2.923(-3) & 6.311(-3) & 8.109(-3) & ... \\
\hline
 & & $E_1$ & $E_2$ & $E_3$ & $E_4$ & $E_5$ & $E_6$ & $E_7$ \\
\hline
Kr$^+$ & 2 & 1.457(3) & 1.204(4) & 5.646(4) & 2.251(5) & 1.616(6) & 5.895(6) & 2.023(7) \\
Kr$^{2+}$ & 1 & 8.328(3) & 1.205(5) & 2.419(6) & 9.715(6) & 4.001(7) & ... & ... \\
 & 2 & 3.952(3) & 4.360(4) & 3.018(5) & 2.593(6) & 9.801(6) & 3.972(7) & ... \\
 & 3 & 4.039(3) & 4.363(4) & 3.080(5) & 2.601(6) & 9.792(6) & 4.040(7) & ... \\
 & 4 & 4.199(3) & 4.438(4) & 3.061(5) & 2.598(6) & 9.797(6) & 3.943(7) & ... \\
 & 5 & 4.136(3) & 4.349(4) & 2.998(5) & 2.602(6) & 9.914(6) & 4.079(7) & ... \\
Kr$^{3+}$ & 1 & 1.864(4) & 1.761(5) & 1.532(6) & 5.351(6) & 2.091(7) & 8.752(7) & ... \\
 & 2 & 7.986(3) & 8.477(4) & 1.324(6) & 5.152(6) & 2.067(7) & 8.783(7) & ... \\
 & 3 & 7.950(3) & 8.445(4) & 1.318(6) & 5.125(6) & 2.029(7) & 8.302(7) & ... \\
 & 4 & 2.237(4) & 1.773(5) & 1.524(6) & 5.352(6) & 2.090(7) & 8.649(7) & ... \\
 & 5 & 2.252(4) & 1.755(5) & 1.520(6) & 5.341(6) & 2.087(7) & 8.585(7) & ... \\
Kr$^{4+}$ & 2 & 3.373(4) & 4.986(5) & 2.372(6) & 8.864(6) & 3.562(7) & 1.567(8) & ... \\
 & 3 & 3.463(4) & 4.926(5) & 2.348(6) & 8.542(6) & 3.413(7) & 1.455(8) & ... \\
 & 4 & 3.585(4) & 4.598(5) & 2.317(6) & 8.722(6) & 3.510(7) & 1.516(8) & ... \\
\hline
\end{tabular}
\tablefoot{Coefficients $c_{\rm i}$ are in cm$^3$\,s$^{-1}$\,K$^{3/2}$, and $E_{\rm i}$ are in K.  States whose RR rate coefficients are accurately described by Equations~\ref{rr1} and \ref{rr2} (with coefficients given in Table~\ref{rrlowtfits}) over the full temperature range (10$^1-10^7$)$z^2$~K are not listed here.  The fits are accurate to within 3\% for Kr$^+$ level~2, and $\leq$2\% otherwise.  The notation $x(y)$ denotes $x\times 10^y$.
\tablefoottext{a}{Note that the ion and level numbers correspond to the target ion (i.e., before recombination).}
}
\end{table*}

\begin{figure}
  \resizebox{\hsize}{!}{\includegraphics{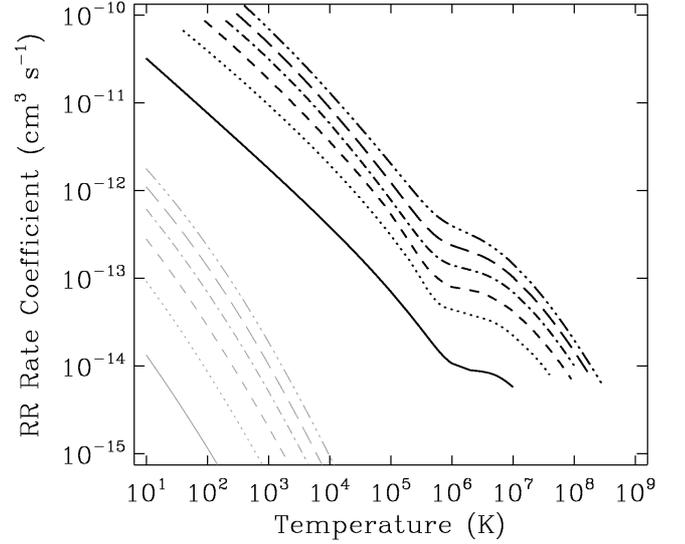}}
  \caption{Radiative recombination rate coefficients (black lines) as a function of temperature for ground state target ions Kr$^+$ (solid curve), Kr$^{2+}$ (dotted curve), Kr$^{3+}$ (short-dashed curve), Kr$^{4+}$ (dash-dot-dash curve), Kr$^{5+}$ (long-dashed curve), and Kr$^{6+}$ (dash-dot-dot-dot-dash curve).  For comparison, the approximate rate coefficients of \citet{mattioli06} are depicted as gray curves, with the line styles corresponding to the same ions.} \label{rrfig}
\end{figure}

We illustrate the RR rate coefficients in Fig.~\ref{rrfig}.  For comparison, we plot the approximate RR rate coefficients of \citet{mattioli06} as gray curves.  It can be seen that the \citet{mattioli06} rate coefficients are at least three orders of magnitude too small over the full range of temperatures.  As stated in their paper, their RR rate coefficients are rough estimates, with valence shell states treated separately from excited shells (which were taken to be hydrogenic).  In the absence of detailed calculations such an approach is reasonable, but the comparison to our results illustrates the dangers of such approximations for heavy element ions.

\subsubsection{Estimation of uncertainties} \label{rr_error}

We have estimated the internal uncertainties of our calculated RR rate coefficients near $T=10^4$~K, the canonical temperature of photoionized astrophysical nebulae.  The adopted CI expansion is again the largest source of uncertainty.  The small and large configuration sets produce RR rate coefficients different from those of the medium CI expansion by 40-45\% for the Kr$^+$ target, and by less than 5-7\% for other species.

We also computed the rate coefficients using a larger number of interpolation points per decade (five, instead of three), using scaling parameters for the target ion (rather than those of the $(N+1)$-electron ion), and without forcing Schmidt orthogonalization of the radial wavefunctions.  These tests resulted in uncertainties no larger than 5\%, except for the two lowest-charge species.  For the Kr$^{2+}$ target, using target ion scaling parameters resulted in rate coefficients differing by up to 10\% for metastable states.  For Kr$^+$ forming Kr$^0$, calculations with target ion scaling parameters led to differences of 20--30\%, while the use of non-orthogonal orbitals resulted in differences of 7\% for the metastable state rate coefficient.

Finally, the $\kappa$-averaged relativistic orbitals used in our calculations prevent AUTOSTRUCTURE from switching to acceleration gauge at high energies.  Thus, the calculated RR rate coefficients at high temperatures (which correspond to PI cross sections at high energies) may be inaccurate.  To determine the temperatures at which such inaccuracies arise, we computed RR rate coefficients using PI cross sections that were terminated at the maximal velocity gauge energy (above which it is necessary to force velocity gauge in AUTOSTRUCTURE).  The rate coefficients calculated in this way exhibited negligible differences with our presented results until $T>10^5$ for the first two ionization stages, or $>$10$^6$~K for higher ions.  This demonstrates that forcing velocity gauge at high energies does not affect our results except at temperatures well above the photoionized regime.

Based on these tests, we estimate the internal uncertainties of the computed RR rate coefficients to be $\leq$10\% except for the Kr$^+$ target, which is uncertain by 50-60\%.

\subsection{Dielectronic recombination rate coefficients}\label{dr}

We determined total and partial final-state resolved DR rate coefficients from the resonant portion of the PI cross sections, with the aid of the ADASDR post-processing code.  The resulting ADAS \textit{adf09} output files \citep[see][]{summers05} are available at the CDS.

The rate coefficients were fit with Equation~\ref{dreq}, with $n$ between 3 and 7, and the fit coefficients are listed in Table~\ref{drfits}.  The non-linear least squares fit algorithm produced fits accurate to within 5\% over the temperature range (10$^1-10^7$)$z^2$~K for most target states, with exceptions noted in Table~\ref{drfits}.  These fits also correctly produce the asymptotic behavior of the rate coefficients outside of this temperature range.

\begin{table*}
\centering
\caption{Fit coefficients for dielectronic recombination rate coefficients; see Equation~\ref{dreq}.}\label{drfits} 
\begin{tabular}{lcccccccc}
\hline \hline
Target & &  &  &  &  &  &  &  \\
Ion\tablefootmark{a} & Level & $c_1$ & $c_2$ & $c_3$ & $c_4$ & $c_5$ & $c_6$ & $c_7$ \\
\hline
Kr$^+$ & 1 & 2.702(-8) & 3.157(-8) & 5.703(-3) & 1.973(-3) & -2.490(-4) & ... & ... \\
 & 2\tablefootmark{b} & 3.343(-3) & 5.611(-3) & -2.453(-4) & ... & ... & ... & ... \\
Kr$^{2+}$ & 1\tablefootmark{c} & 4.418(-7) & 3.859(-7) & 1.710(-6) & 1.270(-5) & 2.232(-4) & 1.101(-2) & ... \\
 & 2\tablefootmark{d} & 1.048(-7) & 1.151(-6) & 8.132(-6) & 8.855(-3) & 7.639(-4) & ... & ... \\
 & 3 & 1.519(-8) & 1.772(-8) & 1.685(-6) & 2.969(-6) & 9.849(-4) & 5.178(-3) & 5.186(-3) \\
 & 4\tablefootmark{b} & 2.047(-7) & 1.263(-6) & 2.406(-3) & 8.124(-3) & -5.667(-4) & ... & ... \\
 & 5\tablefootmark{b} & 3.108(-8) & 8.740(-7) & 6.519(-3) & 9.146(-3) & -3.857(-4) & ... & ... \\
Kr$^{3+}$ & 1 & 6.429(-8) & 1.785(-6) & 1.192(-5) & 6.023(-5) & 1.768(-3) & 3.794(-2) & ... \\
 & 2 & 2.953(-7) & 3.855(-6) & 5.377(-6) & 3.890(-5) & 1.201(-3) & 4.890(-3) & ... \\
 & 3 & 7.020(-7) & 1.569(-6) & 8.266(-6) & 5.250(-5) & 1.946(-3) & 4.908(-3) & ... \\
 & 4 & 1.810(-7) & 4.476(-6) & 6.179(-6) & 6.232(-5) & 1.195(-3) & 3.248(-3) & 3.929(-6) \\
 & 5\tablefootmark{e} & 3.285(-7) & 6.565(-7) & 8.665(-6) & 2.547(-5) & 8.626(-4) & 3.600(-3) & ... \\
Kr$^{4+}$ & 1 & 2.405(-5) & 3.564(-4) & 3.521(-4) & 1.882(-3) & 2.860(-2) & ... & ... \\
 & 2\tablefootmark{e} & 2.211(-5) & 1.062(-4) & 1.637(-4) & 3.685(-4) & 5.024(-3) & 1.617(-2) & ... \\
 & 3 \tablefootmark{e}& 2.211(-5) & 1.062(-4) & 1.637(-4) & 5.925(-4) & 1.199(-2) & 8.986(-3) & ... \\
 & 4 & 1.749(-6) & 2.143(-5) & 8.714(-5) & 5.601(-4) & 1.032(-2) & 8.649(-4) & ... \\
 & 5 & 2.919(-6) & 1.430(-6) & 3.102(-5) & 4.766(-4) & 4.236(-3) & 2.057(-3) & ... \\
Kr$^{5+}$ & 1 & 2.835(-5) & 1.697(-4) & 4.897(-4) & 6.692(-3) & 5.204(-2) & ... & ... \\
 & 2 & 1.795(-5) & 7.020(-5) & 2.474(-4) & 2.299(-3) & 3.064(-2) & 7.897(-4) & ... \\
Kr$^{6+}$ & 1 & 4.030(-5) & 3.483(-4) & 8.082(-4) & 3.569(-3) & 5.891(-2) & 5.091(-3) & ... \\
\hline
 & & $E_1$ & $E_2$ & $E_3$ & $E_4$ & $E_5$ & $E_6$ & $E_7$ \\
\hline
Kr$^+$ & 1 & 5.892 & 4.315(3) & 1.879(5) & 3.015(5) & 6.041(5) & ... & ... \\
 & 2\tablefootmark{b} & 1.667(5) & 2.254(5) & 4.144(5) & ... & ... & ... & ... \\
Kr$^{2+}$ & 1\tablefootmark{c} & 1.291(2) & 3.833(2) & 2.981(3) & 1.509(4) & 1.020(5) & 2.551(5) & ... \\
 & 2\tablefootmark{d} & 4.354(2) & 3.535(3) & 1.311(4) & 2.215(5) & 3.538(5) & ... & ... \\
 & 3 & 4.857(2) & 7.835(2) & 3.365(3) & 1.129(4) & 1.619(5) & 2.507(5) & 2.542(5) \\
 & 4\tablefootmark{b} & 1.423(3) & 1.884(4) & 1.594(5) & 2.716(5) & 5.653(5) & ... & ... \\
 & 5\tablefootmark{b} & 1.845(4) & 2.760(4) & 1.645(5) & 2.651(5) & 5.436(5) & ... & ... \\
Kr$^{3+}$ & 1 & 8.804(2) & 2.124(3) & 7.055(3) & 2.811(4) & 1.382(5) & 2.841(5) & ... \\
 & 2 & 5.889(2) & 1.695(3) & 5.920(3) & 2.667(4) & 1.303(5) & 2.500(5) & ... \\
 & 3 & 1.086(2) & 1.203(3) & 5.564(3) & 2.896(4) & 1.508(5) & 2.616(5) & ... \\
 & 4 & 5.004(2) & 2.294(3) & 6.472(3) & 3.118(4) & 1.102(5) & 2.250(5) & 2.075(6) \\
 & 5\tablefootmark{e} & 1.086(3) & 2.987(3) & 7.146(3) & 2.359(4) & 9.189(4) & 2.156(5) & ... \\
Kr$^{4+}$ & 1 & 6.805(2) & 2.709(3) & 8.633(3) & 1.065(5) & 2.845(5) & ... & ... \\
 & 2\tablefootmark{e} & 6.057(2) & 1.977(3) & 1.026(4) & 4.609(4) & 1.721(5) & 2.851(5) & ... \\
 & 3\tablefootmark{e} & 6.057(2) & 1.977(3) & 1.026(4) & 5.322(4) & 2.198(5) & 3.150(5) & ... \\
 & 4 & 1.640(2) & 1.977(3) & 1.059(4) & 5.558(4) & 2.119(5) & 4.788(5) & ... \\
 & 5 & 2.779(2) & 1.215(3) & 1.378(4) & 5.925(4) & 1.594(5) & 2.687(5) & ... \\
Kr$^{5+}$ & 1 & 5.883(2) & 3.809(3) & 1.701(4) & 1.317(5) & 2.923(5) & ... & ... \\
 & 2 & 4.553(2) & 3.363(3) & 1.595(4) & 7.402(4) & 2.467(5) & 6.872(5) & ... \\
Kr$^{6+}$ & 1 & 7.036(3) & 1.119(4) & 2.735(4) & 1.026(5) & 2.276(5) & 3.660(5) & ... \\
\hline
\end{tabular}
\tablefoot{  \scriptsize Coefficients $c_{\rm i}$ are in cm$^3$\,s$^{-1}$\,K$^{3/2}$, and $E_{\rm i}$ are in K.  The fits are accurate to within 5\% over the temperature range (10$^1-10^7$)$z^2$~K, unless otherwise noted.  The notation $x(y)$ denotes $x^y$.
\tablefoottext{a}{Note that the ion and level numbers correspond to the target ion (i.e., before recombination).}
\tablefoottext{b}{The least-squares fitting algorithm was not able to accurately fit the DR rate coefficient at all temperatures for level~2 of the Kr$^+$ target, and levels~4 and 5 of Kr$^{2+}$.  We chose fits that minimize fit errors near photoionized plasma temperatures.  For Kr$^+$, the fit is accurate to within 9\% at 20\,000~K, and is not valid for temperatures below 10\,000~K (where the rate coefficient is vanishingly small).  For Kr$^{2+}$, the fits are accurate to within 20\% (level~4) and 11\% (level~5) at $T=40$\,000~K.  The fit for level~5 is not valid below 4\,000~K, where the rate coefficient is extremely small.  The fits are accurate to within 2\% at all other temperatures.}
\tablefoottext{c}{The fit is accurate to within 8\%.}
\tablefoottext{d}{The fit is accurate to within 6\%, and is not valid for temperatures below 200~K.}
\tablefoottext{e}{The fit is accurate to within 7\% below 320~K.}
}
\end{table*}
\normalsize

\begin{figure}
  \resizebox{\hsize}{!}{\includegraphics{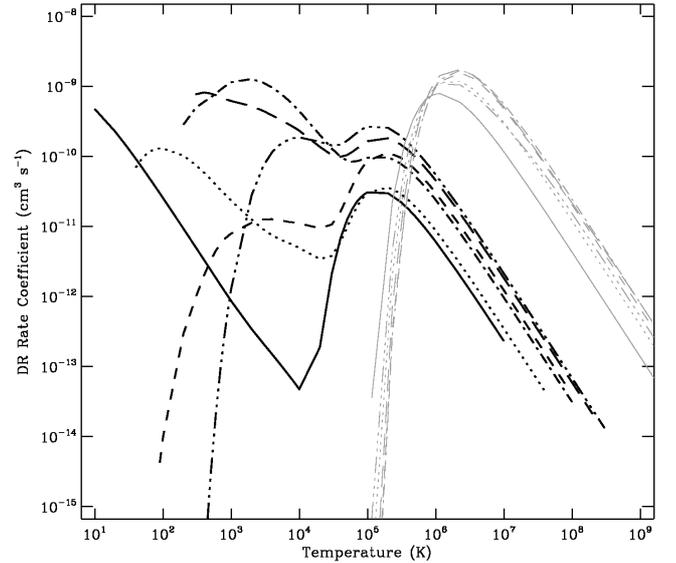}}
  \caption{Dielectronic recombination rate coefficients (black lines) as a function of temperature for ground state target ions Kr$^+$ (solid curve), Kr$^{2+}$ (dotted curve), Kr$^{3+}$ (short-dashed curve), Kr$^{4+}$ (dash-dot-dash curve), Kr$^{5+}$ (long-dashed curve), and Kr$^{6+}$ (dash-dot-dot-dot-dash curve).  DR rate coefficients computed with the Burgess formula over $\Delta n=0$ and $\Delta n=1$ resonances,  from \citet{mattioli06},  are depicted as gray curves, with the line styles corresponding to the same ions.} \label{drfig}
\end{figure}

The DR rate coefficients are plotted in Fig.~\ref{drfig}.  The rate coefficients for different ions are wildly dissimilar at low temperatures due to differences in the structure of near-threshold autoionizing resonances, but show similar behavior at high temperatures.  

\citet{mattioli06} also computed DR rate coefficients for low-charge Kr ions (shown in Fig.~\ref{drfig} as gray curves), in order to properly model its cooling rate coefficient in magnetically-confined fusion plasmas.  These calculations utilized the Burgess formula \citep{burgess65} as modified by \citet{ merts76}, summed over $\Delta n=0$ and 1 resonances.  This formula is well-known to inadequately describe low-temperature DR \citep[e.g., ][]{savin99, savin03}, since the captured electrons are assumed to have thermal energies comparable to the excitation energies of the first resonance transitions of the target ion \citep{mattioli06}.  While this approximation is adequate for collisionally-ionized plasmas, it is not sufficient for photoionized plasmas, in which the electrons are captured into low-energy autoionizing states.  Given this, it is no surprise that the DR rate coefficients from the Burgess formula \citep{mattioli06} are vanishingly small below 10$^5$~K.  At high temperatures ($\geq$ a few times $10^6$~K), the rate coefficients from the Burgess formula exceed those from our calculations, likely because we ignored $\Delta n=1$ core excitations (except in limited cases for Kr$^+$ and Kr$^{2+}$).

\begin{figure}
  \resizebox{\hsize}{!}{\includegraphics{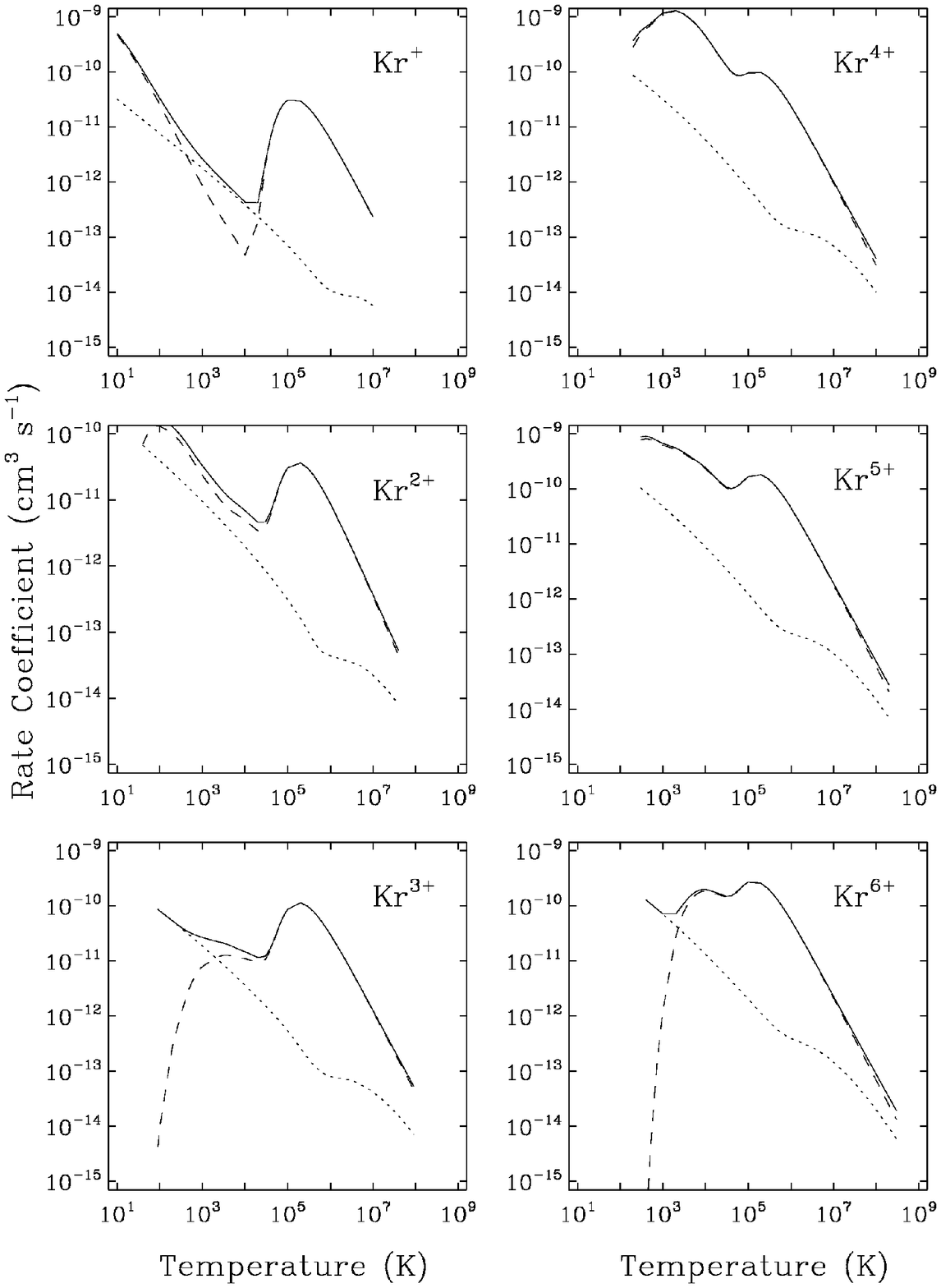}}
  \caption{Comparison of the radiative (dotted curves), dielectronic (dashed curves), and total recombination rate coefficients (solid curves) for the ground state of the first six Kr ions.} \label{totfig}
\end{figure}

Fig.~\ref{totfig} compares the DR (dashed lines), RR (dotted lines), and total recombination rate coefficients (solid lines) for each Kr ion we considered.  As is true of low-charge Se ions \citep{sterling11b}, DR dominates RR for Kr ions near 10$^4$~K, with the DR rate coefficient exceeding that of RR by factors of 2--3 up to two orders of magnitude.  RR is (occasionally) more important than DR only at low temperatures ($\leq 1$--2$\times 10^3$~K).  Only in the case of Kr$^+$ is RR the most important recombination mechanism near $\sim$10$^4$~K.  These results underline the importance of DR in the ionization balance of trans-iron elements.

\subsubsection{Estimation of uncertainties}

DR calculations are inherently more uncertain than those of  PI or RR, since the unknown energies of low-lying autoionizing levels usually dominate other sources of uncertainty.  The energies of these near-threshold autoionizing states have not been experimentally determined for elements beyond the second row of the Periodic Table, and cannot be determined with sufficient accuracy by even the most sophisticated theoretical methods.  To help assuage this uncertainty, we have utilized experimental target energies in all of our DR calculations.

The uncertainty in the DR rate coefficients can be estimated by shifting the continuum relative to near-threshold resonances.  We accomplished this with the ADASDR post-processing code, determining the magnitude of the shift to be the maximum disparity between calculated and experimental target energies, and the direction of the shift by whether the calculated energies were too large or too small.

We find that the uncertainties in the rate coefficients near 10$^4$~K for the ground (metastable) states are 75\% (factor of 20) for Kr$^+$, a factor of 3 (10\% to an order of magnitude) for Kr$^{2+}$, 10\% (15--35\%) for Kr$^{3+}$, 70\% (50--70\%) for Kr$^{4+}$, 50\% (25\%) for Kr$^{5+}$, and a factor of 2 for Kr$^{6+}$.  In contrast, DR calculations with the small CI expansions\footnote{We did not attempt to compute DR rate coefficients with the large CI expansions due to the prohibitive computional requirements.} led to rate coefficients different from those computed with the medium configuration sets by a factor of 3 for Kr$^+$, 60\% for Kr$^{2+}$, a factor of 2 for Kr$^{5+}$, and 10--30\% for other Kr ions.

With the exception of Kr$^{3+}$, for which the uncertainty is probably an underestimate, the DR rate coefficients are uncertain by factors of 2--3 for the ground states of these ions.  These uncertainties are considerably larger than those typically found for our PI cross sections (30--50\%) or RR ($<10$\%).  Given that DR is a much more important recombination mechanism than RR for most Kr ions at photoionized plasma temperatures (Fig.~\ref{totfig}), these uncertainties are significant.  The best hope for accurate low-temperature DR rate coefficients for low-charge ions lies with experimental measurements made with heavy ion storage rings \citep{schippers10}, particularly the forthcoming Cryogenic Storage Ring facility \citep{wolf06}.  Our planned ionization balance calculations for Se and Kr will take into account uncertainties in the atomic data, demonstrating which Se and Kr ions most critically require experimental DR rate coefficients to enable the most accurate possible nebular abundance determinations.


\section{Summary}\label{summary}

We have presented MCBP calculations of distorted-wave photoionization cross sections over the photon energy range 0--100~Ryd, and radiative and dielectronic recombination rate coefficients at temperatures (10$^1-10^7$)$z^2$~K for the first six Kr ions.  The results of all of our calculations are available at the CDS, as well as at http://www.pa.msu.edu/astro/atomicdata/kr\_data.tar.gz.  The RR and DR rate coefficients have been fit to analytical functions, with fit coefficients given in Tables~\ref{rrlowtfits}, \ref{rrhitfits}, and \ref{drfits}.

Kr has been detected in the spectra of planetary nebulae and H~II regions, and its abundance is a sensitive tracer of enrichments by \emph{n}-capture nucleosynthesis.  But in order to derive elemental Kr abundances from those of the observed ions, unseen ionization stages must be accounted for via ionization equilibrium solutions.  Our atomic data determinations, along with charge transfer rate coefficients presented in the third paper of this series \citep{sterling11c}, enable much more accurate Kr ionization balance computations than previously possible.

This study represents the first comprehensive investigation of the photoionization and recombination properties of low-charge Kr ions specifically focused on the photoionized plasma regime.  As such, we have taken significant efforts to provide realistic internal uncertainties for our results, both through comparison to experimental PI measurements, and by testing the sensitivity of our data to changes in the CI expansion and orbital radial scaling parameter values.  

We find typical uncertainties of 30--50\% for our computed PI cross sections, though comparison to experimental measurements indicates that the Kr$^0$ and Kr$^{3+}$ cross sections have larger uncertainties of roughly a factor of two near their ionization thresholds.  We note that our Kr$^+$ PI calculations were significantly improved after comparison with experimental measurements, from accuracies of a factor of 2.5 near the ground state threshold to within $\sim$40\%.  This demonstrates the utility and importance of experimental absolute PI cross-section measurements for constraining and guiding theoretical calculations.

Our RR rate coefficients are more precisely determined, with uncertainties less than 10\% except for the Kr$^+$ target (50--60\%).  On the other hand, the unknown energies of low-lying autoionizing states lead to significantly larger uncertainties in DR rate coefficients, typically factors of 2--3.  These uncertainties are magnified by the fact that DR is the dominant form of recombination for low-charge Kr ions at the temperatures of photoionized astrophysical nebulae, with the rate coefficients exceeding those of RR by as much as two orders of magnitude.  The dominance of DR over RR in photoionized nebulae is common for heavy element ions \citep[e.g.,][]{badnell06a, altun07, nikolic10, sterling11b}.

The estimated uncertainties in these atomic data will play an important role in our forthcoming study of the Se and Kr ionization balance in planetary nebulae.  We will numerically simulate the Se and Kr ionization balance for a wide range of nebular parameters to derive new analytical corrections for unobserved ions.  Using our estimated uncertainties in the PI cross sections and RR and DR rate coefficients, we will run Monte Carlo simulations to test how sensitive elemental abundance determinations are to these uncertainties, thereby illuminating the atomic processes and ionic systems that require additional theoretical and/or experimental investigation.

\acknowledgements

I acknowledge support from an NSF Astronomy and Astrophysics Postdoctoral Fellowship under award AST-0901432 and from NASA grant 06-APRA206-0049.  I am grateful to R.\ Phaneuf, who provided experimental data for Kr$^{3+}$, Kr$^{4+}$, and Kr$^{5+}$ photoionization cross sections; to J.-M.\ Bizau, who kindly sent experimental data for Kr$^+$; and to M.\ Richter and A.\ Sorokin, who provided an analytical fit to the experimental Kr$^0$ PI cross section.  I also thank M.\ Witthoeft and N.\ R.\ Badnell for their careful reading of this manuscript and helpful suggestions that improved its quality.


\bibliographystyle{aa}

\bibliography{sterling_kr.bib}

\clearpage 

\onltab{1}{
\begin{table}
\centering
\caption{Configuration-interaction expansions used for each Kr ion.}\label{ciexp} 
\begin{tabular}{lll}
\hline \hline
Ion & Config.\ Set\tablefootmark{a} & Configurations \\
\hline
Kr$^0$ & Small (13) & 4s$^2$\,4p$^6$,~~4s$^2$\,4p$^5$\,4d \\
 & Medium (391) & [Small]~+~4s$^2$\,4p$^5$\,5s,~~4s$^2$\,4p$^5$\,5p,~~4s$^2$\,4p$^4$\,4d\,5s,~~4s$^2$\,4p$^4$\,4d\,5p \\
 & & 4s\,4p$^6$\,5s,~~4s\,4p$^5$\,4d\,5p,~~3d$^9$\,4s$^2$\,4p$^6$\,4d \\
 & Large (768) & [Medium]~+~4s$^2$\,4p$^5$\,5d,~~4s$^2$\,4p$^4$\,4d$^2$,~~4s$^2$\,4p$^4$\,4d\,5d,~~4s$^2$\,4p$^4$\,5s\,5p \\
 & & 4s\,4p$^5$\,4d$^2$,~~4s\,4p$^5$\,4d\,5s \\
\hline
Kr$^+$ & Small (31) & 4s$^2$\,4p$^5$,~~4s$^2$\,4p$^4$\,4d,~~4s\,4p$^6$ \\
 & Medium (224) & [Small]~+~4s$^2$\,4p$^4$\,5s,~~4s$^2$\,4p$^4$\,5p,~~4s$^2$\,4p$^3$\,4d$^2$,~~4s\,4p$^5$\,4d \\
 & Large (753) & [Medium]~+~4s$^2$\,4p$^4$\,5d,~~4s$^2$\,4p$^3$\,5p$^2$,~~4s\,4p$^5$\,5s,~~4s\,4p$^5$\,5p \\
 & & 4s\,4p$^4$\,4d$^2$,~~4s\,4p$^4$\,4d\,5s,~~3d$^9$\,4s$^2$\,4p$^5$\,4d \\
\hline
Kr$^{2+}$ & Small (10) & 4s$^2$\,4p$^4$,~~4s\,4p$^5$,~~4p$^6$ \\
 & Medium (48) & [Small]~+~4s$^2$\,4p$^3$\,4d \\
 & Large (595) & [Medium]~+~4s$^2$\,4p$^3$\,4f,~~4s$^2$\,4p$^3$\,5s,~~4s$^2$\,4p$^3$\,5p,~~4s$^2$\,4p$^3$\,5d \\
 & & 4s$^2$\,4p$^2$\,4d$^2$,~~4s\,4p$^4$\,4d,~~4s\,4p$^4$\,5s,~~4s\,4p$^4$\,4d\,5s,~~4s\,4p$^4$\,5s\,5p \\
\hline
Kr$^{3+}$ & Small (15) & 4s$^2$\,4p$^3$,~~4s\,4p$^4$,~~4p$^5$ \\
 & Medium (88) & [Small]~+~4s$^2$\,4p$^2$\,4d,~~4s$^2$\,4p\,4d$^2$ \\
 & Large (560) & [Medium]~+~4s$^2$\,4p$^2$\,5s,~~4s$^2$\,4p$^2$\,5p,~~4s$^2$\,4p$^2$\,5d,~~4s$^2$\,4p\,4d\,5d \\
 & & 4s\,4p$^3$\,4d,~~4s\,4p$^2$\,4d$^2$,~~4p$^4$\,4d,~~4p$^4$\,5s \\
\hline
Kr$^{4+}$ & Small (32) & 4s$^2$\,4p$^2$,~~4s$^2$\,4p\,4d,~~4s\,4p$^3$,~~4p$^4$ \\
 & Medium (97) & [Small]~+~4s$^2$\,4d$^2$,~~4s\,4p$^2$\,4d \\
 & Large (765) & [Medium]~+~4s$^2$\,4p\,5s,~~4s$^2$\,4p\,5p,~~4s$^2$\,4p\,5d,~~4s$^2$\,4d\,5d \\
 & & 4s\,4p$^2$\,4f,~~4s\,4p\,4d$^2$,~~4p$^3$\,4d,~~4p$^3$\,5s,~~3d$^9$\,4s$^2$\,4p$^2$\,4d \\
 & & 3d$^9$\,4s$^2$\,4p\,4d\,5s \\
\hline
Kr$^{5+}$ & Small (40) & 4s$^2$\,4p,~~4s$^2$\,4d,~~4s\,4p$^2$,~~4s\,4p\,4d,~~4p$^3$ \\
 & Medium (180) & [Small]~+~4s\,4d$^2$,~~4p$^2$\,4d,~~3d$^9$\,4s$^2$\,4p\,4d \\
 & Large (546) & [Medium]~+~4s$^2$\,5s,~~4s\,4p\,4f,~~4s\,4p\,5s,~~4s\,4p\,5p \\
 & & 4s\,4p\,5d,~~4s\,4d\,5s,~~4s\,4d\,5p,~~4p$^2$\,4d,~~4p$^2$\,5p \\
 & & 4p\,4d$^2$,~~4p\,4d\,5s,~~3d$^9$\,4s$^2$\,4d$^2$,~~3d$^9$\,4s$^2$\,4d\,5s,~~3d$^9$\,4s\,4p$^3$ \\
\hline
Kr$^{6+}$ & Small (10) & 4s$^2$,~~4s\,4p,~~4p$^2$ \\
 & Medium (26) & [Small]~+~4s\,4d,~~4p\,4d \\
 & Large (421) & [Medium]~+~4s\,5s,~~4s\,5p,~~4s\,5d,~~4p\,5s \\
 & & 4p\,5p,~~4d$^2$,~~4d\,5s,~~5s$^2$,~~5p$^2$,~~3d$^9$\,4s$^2$\,4d \\
 & & 3d$^9$\,4s$^2$\,5d,~~3d$^9$\,4s\,4p$^2$,~~3d$^9$\,4s\,4p\,4d,~~3d$^9$\,4s\,5s\,5d \\
\hline
\end{tabular}
\tablefoot{
\tablefoottext{a}{After the name of each configuration set, the number of levels is listed in parentheses.}
}
\end{table}
}

\onllongtab{2}{
\centering
\begin{longtable}{lccc}
\caption{Radial scaling parameters used for CI expansions of each ion}\label{lambdas} \\
\hline \hline
\multicolumn{4}{c}{Kr$^0$} \\
\hline
Orbital & Small & Medium & Large \\
\hline
\endfirsthead
\caption{Continued.} \\
\hline
\endhead
\endfoot
\hline
\endlastfoot
1s & 1.24581 & 1.24943 & 1.25376 \\
2s & 1.09505 & 1.09479 & 1.09436 \\
2p & 1.06562 & 1.06454 & 1.06394 \\
3s & 1.03035 & 1.03063 & 1.03126 \\
3p & 1.01720 & 1.01749 & 1.01734 \\
3d & 0.99378 & 0.99418 & 0.99421 \\
4s & 0.97336 & 0.97642 & 0.97649 \\
4p & 0.97745 & 0.98079 & 0.98061 \\
4d & 0.99528 & 0.99066 & 1.00507 \\
5s & ... & 0.98284 & 0.98945 \\
5p & ... & 1.03498 & 1.03116 \\
5d & ... & ... & 0.98293 \\
\hline
\multicolumn{4}{c}{Kr$^+$} \\
\hline
Orbital & Small & Medium & Large \\
\hline
1s & 1.23879 & 1.25198 & 1.24043 \\
2s & 1.09415 & 1.09360 & 1.09380 \\
2p & 1.06448 & 1.06385 & 1.06579 \\
3s & 1.03091 & 1.03047 & 1.03142 \\
3p & 1.01758 & 1.01736 & 1.01935 \\
3d & 0.99294 & 0.99292 & 0.99332 \\
4s & 0.98148 & 0.98065 & 0.98377 \\
4p & 0.98763 & 0.98679 & 0.99848 \\
4d & 1.00279 & 1.00271 & 1.01442 \\
5s & ... & 0.99546 & 1.00878 \\
5p & ... & 0.99218 & 1.02685 \\
5d & ... & ... & 0.98887 \\
\hline
\multicolumn{4}{c}{Kr$^{2+}$} \\
\hline
Orbital & Small & Medium & Large \\
\hline
1s & 1.24899 & 1.24674 & 1.24682 \\
2s & 1.09401 & 1.09269 & 1.09447 \\
2p & 1.06441 & 1.06369 & 1.06537 \\
3s & 1.03157 & 1.03153 & 1.03222 \\
3p & 1.01798 & 1.01829 & 1.01850 \\
3d & 0.99232 & 0.99236 & 0.99290 \\
4s & 0.98775 & 0.98727 & 0.98943 \\
4p & 0.99891 & 1.00031 & 0.99975 \\
4d & ... & 1.00481 & 0.99865 \\
4f & ... & ... & 1.10407 \\
5s & ... & ... & 1.01412 \\
5p & ... & ... & 1.02416 \\
5d & ... & ... & 0.99391 \\
\hline
\multicolumn{4}{c}{Kr$^{3+}$} \\
\hline
Orbital & Small & Medium & Large \\
\hline
1s & 1.24588 & 1.25627 & 1.26022 \\
2s & 1.09374 & 1.09425 & 1.09349 \\
2p & 1.06337 & 1.06369 & 1.06341 \\
3s & 1.03257 & 1.03186 & 1.03138 \\
3p & 1.01690 & 1.01761 & 1.01693 \\
3d & 0.99187 & 0.99230 & 0.99231 \\
4s & 0.99190 & 0.99209 & 0.99176 \\
4p & 0.99608 & 0.99654 & 0.99708 \\
4d & ... & 0.99583 & 1.00162 \\
5s & ... & ... & 0.99337 \\
5p & ... & ... & 1.02470 \\
5d & ... & ... & 0.99612 \\
\hline
\multicolumn{4}{c}{Kr$^{4+}$} \\
\hline
Orbital & Small & Medium & Large \\
\hline
1s & 1.24395 & 1.23583 & 1.24256 \\
2s & 1.09279 & 1.09208 & 1.09285 \\
2p & 1.06260 & 1.06301 & 1.06304 \\
3s & 1.03102 & 1.03082 & 1.03183 \\
3p & 1.01611 & 1.01680 & 1.01773 \\
3d & 0.99150 & 0.99115 & 0.99280 \\
4s & 0.99377 & 0.99297 & 0.99511 \\
4p & 0.99923 & 1.01125 & 1.01144 \\
4d & 0.98202 & 0.99108 & 1.00730 \\
4f & ... & ... & 1.06230 \\
5s & ... & ... & 0.99320 \\
5p & ... & ... & 1.01922 \\
5d & ... & ... & 0.99607 \\
\hline
\multicolumn{4}{c}{Kr$^{5+}$} \\
\hline
Orbital & Small & Medium & Large \\
\hline
1s & 1.24564 & 1.24269 & 1.24262 \\
2s & 1.09419 & 1.09283 & 1.09278 \\
2p & 1.06235 & 1.06342 & 1.06321 \\
3s & 1.03107 & 1.03184 & 1.03230 \\
3p & 1.01432 & 1.01668 & 1.01678 \\
3d & 0.99078 & 0.99232 & 0.99232 \\
4s & 0.99321 & 0.99567 & 0.99592 \\
4p & 0.99644 & 1.00920 & 1.01108 \\
4d & 0.97991 & 1.00907 & 1.00657 \\
4f & ... & ... & 0.99064 \\
5s & ... & ... & 1.01040 \\
5p & ... & ... & 0.99061 \\
5d & ... & ... & 0.98874 \\
\hline
\multicolumn{4}{c}{Kr$^{6+}$} \\
\hline
Orbital & Small & Medium & Large \\
\hline
1s & 1.24521 & 1.25082 & 1.25544 \\
2s & 1.09024 & 1.09064 & 1.09086 \\
2p & 1.06163 & 1.06156 & 1.06148 \\
3s & 1.02882 & 1.02851 & 1.02813 \\
3p & 1.01296 & 1.01303 & 1.01355 \\
3d & 0.98920 & 0.98923 & 0.98945 \\
4s & 0.99077 & 0.99072 & 0.99184 \\
4p & 0.98740 & 0.98739 & 0.98756 \\
4d & ... & 0.97409 & 0.97560 \\
5s & ... & ... & 1.02656 \\
5p & ... & ... & 0.98759 \\
5d & ... & ... & 1.00181 \\
\end{longtable}
}

\renewcommand{\thefootnote}{\alph{footnote}}

\onllongtab{3}{
\centering
\begin{longtable}{ccccccc}
\caption{Comparison of selected calculated and experimental energies (in Rydbergs)}\label{ecomp} \\
\hline \hline
\multicolumn{7}{c}{Kr$^0$} \\
\hline
Index & Config. & Term & Small & Medium & Large & NIST \\
\hline
\endfirsthead
\caption{Continued.} \\
\endhead
\endfoot
\hline
\endlastfoot
1 & 4s$^2$\,4p$^6$ & $^1$S$_0$ & 0.0000 & 0.0000 & 0.0000 & 0.0000 \\
2 & 4s$^2$\,4p$^5$\,5s & $^3$P$_2$ & ... & 0.7389 & 0.7322 & 0.7288 \\
3 &  & $^3$P$_1$ &  ... & 0.7492 & 0.7427 & 0.7374 \\
4 &  & $^3$P$_0$ &  ... & 0.7781 & 0.7711 & 0.7763 \\
5 &  & $^1$P$_1$ &  ... & 0.7868 & 0.7803 & 0.7823 \\
6 & 4s$^2$\,4p$^5$\,5p & $^3$S$_1$ & ... & 0.8301 & 0.8254 & 0.8308 \\
7 &  & $^3$D$_2$ & ... & 0.8482 & 0.8438 & 0.8410 \\
8 &  & $^1$P$_1$ & ... & 0.8567 & 0.8514 & 0.8412 \\
9 &  & $^3$P$_2$ & ... & 0.8605 & 0.8550 & 0.8472 \\
10 &  & $^3$P$_0$ & ... & 0.8833 & 0.8776 & 0.8486 \\
11 &  & $^3$D$_3$ & ... & 0.8489 & 0.8431 & 0.8574 \\
12 & 4s$^2$\,4p$^5$\,4d & $^3$P$_0$ & 0.8573 & 0.8918 & 0.8846 & 0.8818 \\
13 &  & $^3$P$_1$ & 0.8596 & 0.8938 & 0.8869 & 0.8847 \\
14 & 4s$^2$\,4p$^5$\,5p & $^3$D$_1$ & ... & 0.8873 & 0.8822 & 0.8894 \\
15 & 4s$^2$\,4p$^5$\,4d & $^3$F$_4$ & 0.8630 & 0.8959 & 0.8909 & 0.8902 \\
16 & 4s$^2$\,4p$^5$\,5p & $^3$P$_1$ & ... & 0.8940 & 0.8886 & 0.8912 \\
17 & 4s$^2$\,4p$^5$\,4d & $^3$P$_2$ & 0.8637 & 0.8973 & 0.8912 & 0.8925 \\
18 &  & $^3$F$_3$ & 0.8655 & 0.8979 & 0.8935 & 0.8951 \\
19 &  & $^1$D$_2$ & 0.8703 & 0.9027 & 0.8987 & 0.9008 \\
20 &  & $^3$D$_3$ & 0.8715 & 0.9037 & 0.8998 & 0.9009 \\
21 &  & $^3$D$_1$ & 0.8822 & 0.9139 & 0.9113 & 0.9079 \\
22 & 4s$^2$\,4p$^5$\,5p & $^1$D$_2$ & ... & 0.8937 & 0.8883 & 0.9080 \\
I.P. & ... & ... & 0.9915 & 1.0218 & 1.0177 & 1.0290 \\
\hline
\multicolumn{7}{c}{Kr$^+$} \\
\hline
Index & Config. & Term & Small & Medium & Large & NIST \\
\hline
1 & 4s$^2$\,4p$^5$ & $^2$P$_{3/2}$ & 0.0000 & 0.0000 & 0.0000 & 0.0000 \\
2 &  & $^2$P$_{1/2}$ & 0.0448 & 0.0441 & 0.0487 & 0.0489 \\
3 & 4s\,4p$^6$ & $^2$S$_{1/2}$ & 1.0103 & 1.0506 & 1.0213 & 0.9933 \\
4 & 4s$^2$\,4p$^4$\,5s & $^4$P$_{5/2}$ & ... & 1.0437 & 1.0277 & 1.0282 \\
5 &  & $^4$P$_{3/2}$ & ... & 1.0650 & 1.0508 & 1.0488 \\
6 &  & $^4$P$_{1/2}$ & ... & 1.0808 & 1.0682 & 1.0717 \\
7 &  & $^2$P$_{3/2}$ & ... & 1.1014 & 1.0891 & 1.0796 \\
8 & 4s$^2$\,4p$^4$\,4d & $^4$D$_{7/2}$ & 1.0811 & 1.1246 & 1.1011 & 1.0954 \\
9 &  & $^4$D$_{5/2}$ & 1.0833 & 1.1267 & 1.1032 & 1.0974 \\
10 &  & $^4$D$_{3/2}$ & 1.0880 & 1.1314 & 1.1083 & 1.1026 \\
11 & 4s$^2$\,4p$^4$\,5s & $^2$P$_{1/2}$ & ... & 1.1260 & 1.1188 & 1.1027 \\
12 & 4s$^2$\,4p$^4$\,4d  & $^4$D$_{1/2}$ & 1.0930 & 1.1374 & 1.1116 & 1.1097 \\
13 &  & $^4$F$_{9/2}$ & 1.1282 & 1.1717 & 1.1600 & 1.1482 \\
14 & 4s$^2$\,4p$^4$\,5s & $^2$D$_{3/2}$ & ... & 1.1912 & 1.1811 & 1.1628 \\
15 &  & $^2$D$_{5/2}$ & ... & 1.1941 & 1.1848 & 1.1652 \\
16 & 4s$^2$\,4p$^4$\,4d & $^4$F$_{7/2}$ & 1.1437 & 1.1871 & 1.1763 & 1.1658 \\
17 &  & $^4$F$_{5/2}$ & 1.1563 & 1.1997 & 1.1905 & 1.1819 \\
18 &  & $^4$F$_{3/2}$ & 1.1630 & 1.2063 & 1.1980 & 1.1893 \\
19 &  & $^4$P$_{1/2}$ & 1.1642 & 1.2126 & 1.1946 & 1.1928 \\
20 &  & $^2$P$_{3/2}$ & 1.1997 & 1.2490 & 1.2299 & 1.1972 \\
21 &  & $^2$F$_{7/2}$ & 1.1806 & 1.2241 & 1.2137 & 1.1995 \\
22 &  & $^4$P$_{5/2}$ & 1.1846 & 1.2126 & 1.2168 & 1.2117 \\
23 & 4s$^2$\,4p$^4$\,5p & $^4$P$_{5/2}$ & ... & 1.2244 & 1.2344 & 1.2204 \\
24 &  & $^4$P$_{3/2}$ & ... & 1.2289 & 1.2385 & 1.2237 \\
25 & 4s$^2$\,4p$^4$\,4d & $^2$F$_{5/2}$ & 1.2041 & 1.2489 & 1.2402 & 1.2263 \\
26 & 4s$^2$\,4p$^4$\,5p & $^4$P$_{1/2}$ & ... & 1.2126 & 1.2536 & 1.2373 \\
27 &  & $^4$D$_{7/2}$ & ... & 1.2443 & 1.2633 & 1.2373 \\
28 &  & $^4$D$_{5/2}$ & ... & 1.2782 & 1.3002 & 1.2400 \\
29 &  & $^4$D$_{1/2}$ & ... & 1.2781 & 1.2983 & 1.2772 \\
30 &  & $^4$D$_{3/2}$ & ... & 1.2694 & 1.2903 & 1.2610 \\
I.P. & ... & ... & 1.7461 & 1.7636 & 1.7555 & 1.7904 \\
\hline
\multicolumn{7}{c}{Kr$^{2+}$} \\
\hline
Index & Config. & Term & Small & Medium & Large & NIST \\
\hline
1 & 4s$^2$\,4p$^4$ & $^3$P$_2$ & 0.0000 & 0.0000 & 0.0000 & 0.0000 \\
2 &  & $^3$P$_1$ & 0.0386 & 0.0394 & 0.0385 & 0.0414 \\
3 &  & $^3$P$_0$ & 0.0457 & 0.0464 & 0.0463 & 0.0484 \\
4 &  & $^1$D$_2$ & 0.1550 & 0.1558 & 0.1498 & 0.1334 \\
5 &  & $^1$S$_0$ & 0.2850 & 0.2862 & 0.3019 & 0.3014 \\
6 & 4s\,4p$^5$ & $^3$P$_2$ & 1.2254 & 1.0238 & 1.0951 & 1.0564 \\
7 &  & $^3$P$_1$ & 1.2588 & 1.0543 & 1.1255 & 1.0879 \\
8 &  & $^3$P$_0$ & 1.2773 & 1.0723 & 1.1434 & 1.1076 \\
9 & 4s$^2$\,4p$^3$\,4d & $^5$D$_0$ & ... & 1.2089 & 1.2840 & 1.2616 \\
10 &  & $^5$D$_1$ & ... & 1.2091 & 1.2842 & 1.2618 \\
11 &  & $^5$D$_2$ & ... & 1.2093 & 1.2843 & 1.2619 \\
12 &  & $^5$D$_3$ & ... & 1.2097 & 1.2846 & 1.2620 \\
13 &  & $^5$D$_4$ & ... & 1.2110 & 1.2857 & 1.2635 \\
14 &  & $^1$P$_1$ & ... & 1.2834 & 1.3521 & 1.2929 \\
15 &  & $^3$D$_2$ & ... & 1.3246 & 1.3992 & 1.3469 \\
16 &  & $^3$D$_3$ & ... & 1.3300 & 1.4047 & 1.3554 \\
17 &  & $^3$D$_1$ & ... & 1.3306 & 1.4078 & 1.3584 \\
18 &  & $^1$S$_0$ & ... & 1.4054 & 1.4801 & 1.4070 \\
I.P. & ... & ... & 2.5527 & 2.5358 & 2.5644 & 2.7158 \\
\hline
\multicolumn{7}{c}{Kr$^{3+}$} \\
\hline
Index & Config. & Term & Small & Medium & Large & NIST \\
\hline
1 & 4s$^2$\,4p$^3$ & $^4$S$_{3/2}$ & 0.0000 & 0.0000 & 0.0000 & 0.0000 \\
2 &  & $^2$D$_{3/2}$ & 0.1919 & 0.1895 & 0.1851 & 0.1553 \\
3 &  & $^2$D$_{5/2}$ & 0.2076 & 0.2073 & 0.1992 & 0.1704 \\
4 &  & $^2$P$_{1/2}$ & 0.2944 & 0.2780 &  0.2981 & 0.2830 \\
5 &  & $^2$P$_{3/2}$ & 0.3139 & 0.3001 & 0.3162 & 0.3044 \\
6 & 4s\,4p$^4$ & $^4$P$_{5/2}$ & 1.1236 & 1.0476 & 1.0560 & 1.0822 \\
7 &  & $^4$P$_{3/2}$ & 1.1559 & 1.0782 & 1.0862 & 1.1156 \\
8 &  & $^4$P$_{1/2}$ & 1.1718 & 1.0932 & 1.1010 & 1.1310 \\
9 &  & $^2$D$_{3/2}$ & 1.5334 & 1.3520 & 1.3390 & 1.3284 \\
10 &  & $^2$D$_{5/2}$ & 1.5363 & 1.3564 & 1.3438 & 1.3363 \\
11 & 4s$^2$\,4p$^2$\,4d & $^2$P$_{3/2}$ & ... & 1.5231 & 1.5022 & 1.4894 \\
12 &  & $^4$F$_{3/2}$ & ... & 1.5965 & 1.5861 & 1.5740 \\
13 & 4s\,4p$^4$ & $^2$S$_{1/2}$ & 1.7359 & 1.6060 & 1.6041 & 1.5852 \\
14 & 4s$^2$\,4p$^2$\,4d & $^4$F$_{5/2}$ & ... & 1.6075 & 1.5969 & 1.5866 \\
15 &  & $^4$F$_{7/2}$ & ... & 1.6242 & 1.6135 & 1.6059 \\
16 &  & $^4$F$_{9/2}$ & ... & 1.6459 & 1.6355 & 1.6300 \\
17 &  & $^2$F$_{5/2}$ & ... & 1.6912 & 1.6716 & 1.6332 \\
18 &  & $^4$D$_{1/2}$ & ... & 1.6677 & 1.6484 & 1.6465 \\
19 &  & $^4$D$_{7/2}$ & ... & 1.6746 & 1.6554 & 1.6472 \\
20 &  & $^4$D$_{3/2}$ & ... & 1.6705 & 1.6512 & 1.6494 \\
21 &  & $^4$D$_{5/2}$ & ... & 1.6668 & 1.6475 & 1.6646 \\
22 &  & $^2$F$_{7/2}$ & ... & 1.7262 & 1.7061 & 1.7001 \\
23 &  & $^4$P$_{5/2}$ & ... & 1.9196 & 1.8971 & 1.8355 \\
24 &  & $^4$P$_{1/2}$ & ... & 1.9416 & 1.9299 & 1.8657 \\
I.P. & ... & ... & 3.6801 & 3.6416 & 3.6958 & 3.8583 \\
\hline
\multicolumn{7}{c}{Kr$^{4+}$} \\
\hline
Index & Config. & Term & Small & Medium & Large & NIST \\
\hline
1 & 4s$^2$\,4p$^2$ & $^3$P$_0$ & 0.0000 & 0.0000 & 0.0000 & 0.0000 \\
2 &  & $^3$P$_1$ & 0.0315 & 0.0334 & 0.0336 & 0.0341 \\
3 &  & $^3$P$_2$ & 0.0664 & 0.0701 & 0.0702 & 0.0692 \\
4 &  & $^1$D$_2$ & 0.2012 & 0.2049 & 0.2033 & 0.1797 \\
5 &  & $^1$S$_0$ & 0.3428 & 0.3808 & 0.3691 & 0.3573 \\
6 & 4s\,4p$^3$ & $^3$D$_1$ & 1.1403 & 1.2155 & 1.1855 & 1.1815 \\
7 &  & $^3$D$_2$ & 1.1403 & 1.2156 & 1.1859 & 1.1826 \\
8 &  & $^3$D$_3$ & 1.1487 & 1.2252 & 1.1955 & 1.1939 \\
9 &  & $^3$P$_0$ & 1.3188 & 1.3975 & 1.3685 & 1.3480 \\
10 &  & $^3$P$_1$ & 1.3208 & 1.3998 & 1.3706 & 1.3513 \\
11 &  & $^3$P$_2$ & 1.3225 & 1.4016 & 1.3715 & 1.3548 \\
12 & 4s$^2$\,4p\,4d & $^1$D$_2$ & 1.5108 & 1.5844 & 1.5110 & 1.4889 \\
13 & 4s\,4p$^3$ & $^3$S$_1$ & 1.8518 & 1.9392 & 1.7978 & 1.6864 \\
14 & 4s$^2$\,4p\,4d & $^3$F$_2$ & 1.7476 & 1.8119 & 1.7452 & 1.7340 \\
15 &  & $^3$F$_3$ & 1.7695 & 1.8353 & 1.7687 & 1.7583 \\
16 & 4s\,4p$^3$ & $^1$P$_1$ & 1.7866 & 1.8678 & 1.8739 & 1.7682 \\
17 & 4s$^2$\,4p\,4d & $^3$P$_2$ & 1.9900 & 2.0595 & 1.9859 & 1.9258 \\
18 &  & $^3$P$_1$ & 2.0101 & 2.0807 & 2.0060 & 1.9495 \\
19 &  & $^3$P$_0$ & 2.0287 & 2.1016 & 2.0278 & 1.9721 \\
20 & 4s\,4p$^3$ & $^1$D$_2$ & 2.0821 & 2.1628 & 2.0666 & 1.9763 \\
21 & 4s$^2$\,4p\,4d & $^3$D$_1$ & 2.0563 & 2.1292 & 2.0507 & 1.9934 \\
22 &  & $^3$D$_3$ & 2.0681 & 2.1409 & 2.0612 & 1.9992 \\
23 &  & $^3$D$_2$ & 2.0665 & 2.1400 & 2.0578 & 2.0032 \\
24 &  & $^1$P$_1$ & 2.3043 & 2.3824 & 2.2647 & 2.1663 \\
25 &  & $^1$F$_3$ & 2.2196 & 2.2912 & 2.2152 & 2.1955 \\
I.P. & ... & ... & 4.6132 & 4.6721 & 4.6669 & 4.7550 \\
\hline
\multicolumn{7}{c}{Kr$^{5+}$} \\
\hline
Index & Config. & Term & Small & Medium & Large & NIST \\
\hline
1 & 4s$^2$\,4p & $^2$P$_{1/2}$ & 0.0000 & 0.0000 & 0.0000 & 0.0000 \\
2 &  & $^2$P$_{3/2}$ & 0.0678 & 0.0730 & 0.0738 & 0.0739 \\
3 & 4s\,4p$^2$ & $^4$P$_{1/2}$ & 0.9424 & 0.9711 & 0.9831 & 0.9827 \\
4 &  & $^4$P$_{3/2}$ & 0.9692 & 1.0003 & 1.0126 & 1.0133 \\
5 &  & $^4$P$_{5/2}$ & 1.0056 & 1.0387 & 1.0515 & 1.0523 \\
6 &  & $^2$D$_{3/2}$ & 1.3229 & 1.3213 & 1.3276 & 1.2910 \\
7 &  & $^2$D$_{5/2}$ & 1.3279 & 1.3283 & 1.3348 & 1.3006 \\
8 &  & $^2$S$_{1/2}$ & 1.6289 & 1.6268 & 1.6365 & 1.5499 \\
9 &  & $^2$P$_{1/2}$ & 1.7936 & 1.7402 & 1.7513 & 1.6434 \\
10 & & $^2$P$_{3/2}$ & 1.8328 & 1.7761 & 1.7875 & 1.6751 \\
11 & 4s$^2$\,4d & $^2$D$_{3/2}$ & 2.1468 & 2.1161 & 2.0720 & 2.0241 \\
12 &  & $^2$D$_{5/2}$ & 2.1535 & 2.1239 & 2.0797 & 2.0325 \\
13 & 4p$^3$ & $^2$D$_{3/2}$ & 2.4829 & 2.5286 & 2.5213 & 2.5152 \\
14 &  & $^2$D$_{5/2}$ & 2.4947 & 2.5416 & 2.5353 & 2.5339 \\
15 &  & $^4$S$_{3/2}$ & 2.5397 & 2.5959 & 2.5756 & 2.5405 \\
16 &  & $^2$P$_{1/2}$ & 2.7983 & 2.8514 & 2.8225 & 2.7675 \\
17 &  & $^2$P$_{3/2}$ & 2.8045 & 2.8584 & 2.8320 & 2.7829 \\
18 & 4s\,4p\,4d & $^4$P$_{5/2}$ & 2.9948 & 3.0422 & 3.0359 & 3.0250 \\
19 &  & $^4$D$_{3/2}$ & 3.0498 & 3.1014 & 3.0939 & 3.0357 \\
20 &  & $^4$D$_{1/2}$ & 3.0185 & 3.0676 & 3.0597 & 3.0431 \\
21 &  & $^4$P$_{1/2}$ & 3.0422 & 3.0926 & 3.0865 & 3.0804 \\
22 &  & $^4$D$_{7/2}$ & 3.0560 & 3.1098 & 3.0999 & 3.0812 \\
23 &  & $^4$P$_{3/2}$ & 3.0073 & 3.0555 & 3.0488 & 3.0834 \\
24 &  & $^4$D$_{5/2}$ & 3.0549 & 3.1077 & 3.0990 & 3.0842 \\
I.P. & ... & ... & 5.6991 & 5.7388 & 5.7207 & 5.7692 \\
\hline
\multicolumn{7}{c}{Kr$^{6+}$} \\
\hline
Index & Config. & Term & Small & Medium & Large & NIST \\
\hline
1 & 4s$^2$ & $^1$S$_0$ & 0.0000 & 0.0000 & 0.0000 & 0.0000 \\
2 & 4s\,4p & $^3$P$_0$ & 1.0414 & 1.0309 & 1.0589 & 1.0697 \\
3 &  & $^3$P$_1$ & 1.0640 & 1.0532 & 1.0812 & 1.0944 \\
4 &  & $^3$P$_2$ & 1.1155 & 1.1046 & 1.1326 & 1.1532 \\
5 &  & $^1$P$_1$ & 1.6648 & 1.5866 & 1.6082 & 1.5568 \\
6 & 4p$^2$ & $^3$P$_0$ & 2.4935 & 2.4935 & 2.5192 & 2.5054 \\
7 &  & $^3$P$_1$ & 2.5271 & 2.5271 & 2.5538 & 2.5462 \\
8 &  & $^1$D$_2$ & 2.7151 & 2.5133 & 2.5428 & 2.5489 \\
9 &  & $^3$P$_2$ & 2.5677 & 2.5922 & 2.6196 & 2.6262 \\
10 &  & $^1$S$_0$ & 3.0307 & 3.0306 & 2.9919 & 2.9324 \\
11 & 4s\,4d & $^3$D$_1$ & ... & 3.1590 & 3.1913 & 3.1892 \\
12 &  & $^3$D$_2$ & ... & 3.1627 & 3.1951 & 3.1932 \\
13 &  & $^3$D$_3$ & ... & 3.1685 & 3.2010 & 3.1996 \\
14 &  & $^1$D$_2$ & ... & 3.5333 & 3.5344 & 3.4582 \\
15 & 4p\,4d & $^1$D$_2$ & ... & 4.3817 & 4.4200 & 4.4438 \\
I.P.\footnotemark[1] & ... & ... & ... & ... & ... & 8.1586 \\
\footnotetext[1]{Since we did not compute the electronic structure of Kr$^{7+}$, it was not possible to calculate the ionization potential of Kr$^{6+}$.}
\end{longtable}
}

\renewcommand{\thefootnote}{\alph{footnote}}
\onllongtab{4}{
\centering
\begin{longtable}{llcccc}
\caption{Calculated Einstein A-coefficients (in s$^{-1}$) compared to literature values.  The notation $x(y)$ denotes $x\times 10^y$.}\label{acomp} \\
\hline \hline
\multicolumn{6}{c}{Kr$^0$} \\
\hline
$i$ & $j$ & Small & Medium & Large & Previous\footnotemark[1] \\
\hline
\endfirsthead
\caption{Continued.} \\
\hline
\endhead
\hline
\endfoot
\hline
\endlastfoot
1 & 6 & 1.34(7) & 6.41(8) & 5.79(8) & 2.98(8) \\
1 & 5 & ... & 1.02(9) & 9.56(8) & 3.09(8) \\
1 & 13 & ... & 4.55(7) & 3.21(7) & 1.11(7) \\
\hline
\multicolumn{6}{c}{Kr$^+$} \\
\hline
$i$ & $j$ & Small & Medium & Large & Previous\footnotemark[1] \\
\hline
1 & 2 & 2.14 & 2.04 & 2.74 & 2.80 \\
1 & 3 & 5.84(7) & 2.71(7) & 2.02(6) & 2.27(7) \\
2 & 3 & 4.79(7) & 3.87(7) & 1.42(7) & 9.76(6) \\
\hline
\multicolumn{6}{c}{Kr$^{2+}$} \\
\hline
$i$ & $j$ & Small & Medium & Large & Previous\footnotemark[1] \\
\hline
1 & 2 & 1.66 & 1.76 & 1.77 & 2.01 \\
1 & 3 & 1.87(-3) & 2.01(-3) & 2.28(-3) & 2.38(-3) \\
2 & 3 & 2.45(-2) & 2.39(-2) & 3.19(-2) & 2.93(-2) \\
1 & 4 & 5.03 & 5.42 & 4.82 & 4.73 \\
2 & 4 & 7.22(-1) & 7.43(-1) & 6.71(-1) & 5.38(-1) \\
3 & 4 & 1.67(-3) & 1.73(-3) & 8.95(-4) & 2.90(-4) \\
1 & 5 & 1.85(-1) & 1.88(-1) & 3.28(-1) & 6.37(-1) \\
2 & 5 & 4.54(1) & 4.72(1) & 4.82(1) & 5.30(1) \\
4 & 5 & 1.24 & 1.24 & 2.79 & 4.12 \\
\hline
\multicolumn{6}{c}{Kr$^{3+}$} \\
\hline
$i$ & $j$ & Small & Medium & Large & Previous\footnotemark[1] \\
\hline
1 & 2 & 3.34 & 3.94 & 2.75 & 3.14 \\
1 & 3 & 1.51(-1) & 1.67(-1) & 1.31(-1) & 1.52(-1) \\
2 & 3 & 4.84(-2) & 6.82(-2) & 3.58(-2) & 4.81(-2) \\
1 & 4 & 1.01(1) & 9.42 & 1.00(1) & 1.26(1) \\
2 & 4 & 3.36 & 2.65 & 3.83 & 5.35 \\
3 & 4 & 4.42(-2) & 1.57(-2) & 8.91(-2) & 1.36(-1) \\
1 & 5 & 1.93(1) & 1.73(1) & 1.98(1) & 2.35(1) \\
2 & 5 & 8.37 & 7.50 & 8.93 & 1.34(1) \\
3 & 5 & 3.53 & 2.86 & 4.03 & 5.93 \\
4 & 5 & 7.60(-2) & 1.07(-1) & 6.23(-2) & 9.99(-2) \\
\hline
\multicolumn{6}{c}{Kr$^{4+}$} \\
\hline
$i$ & $j$ & Small & Medium & Large & Previous\footnotemark[1] \\
\hline
1 & 2 & 7.22(-1) & 8.57(-1) & 8.74(-1) & 9.26(-1) \\
1 & 3 & 1.68(-3) & 1.97(-3) & 2.08(-3) & 1.90(-3) \\
2 & 3 & 7.15(-1) & 8.35(-1) & 8.24(-1) & 6.90(-1) \\
1 & 4 & 6.80(-3) & 4.46(-3) & 5.69(-3) & 5.53(-4) \\
2 & 4 & 4.37 & 5.26 & 5.21 & 4.98 \\
3 & 4 & 6.27 & 7.22 & 7.11 & 5.99 \\
2 & 5 & 6.72(1) & 8.63(1) & 8.22(1) & 7.95(1) \\
3 & 5 & 7.30(-1) & 1.66 & 1.53 & 1.90 \\
4 & 5 & 1.36 & 3.70 & 2.99 & 3.61 \\
\hline
\multicolumn{6}{c}{Kr$^{5+}$} \\
\hline
$i$ & $j$ & Small & Medium & Large & Previous\footnotemark[1] \\
\hline
1 & 2 & 3.70 & 4.62 & 4.78 & 4.83 \\
1 & 3 & 5.94(6) & 8.69(6) & 9.59(6) & 1.47(7) \\
2 & 3 & 2.84(6) & 3.54(6) & 3.92(6) & 5.95(6) \\
1 & 4 & 9.53(4) & 1.21(5) & 1.57(5) & 2.33(5) \\
2 & 4 & 1.32(6) & 2.11(6) & 2.33(6) & 3.29(6) \\
2 & 5 & 4.59(6) & 7.10(6) & 8.31(6) & 1.27(7) \\
\hline
\multicolumn{6}{c}{Kr$^{6+}$} \\
\hline
$i$ & $j$ & Small & Medium & Large & Previous\footnotemark[1] \\
\hline
1 & 3 & 1.04(7) & 1.31(7) & 1.56(7) & 6.39(7) \\
1 & 5 & 1.30(10) & 1.16(10) & 1.28(10) & 1.60(10) \\
\footnotetext[1]{References for forbidden transitions in the ground configuration: Kr$^+$ \citep{biemont88}, Kr$^{2+}$ \citep{biemont86b}, Kr$^{3+}$ and Kr$^{4+}$ \citep{biemont86a}, Kr$^{5+}$ \citep{biemont87}.  \citet{morton00} is the reference for Kr$^0$, Kr$^{+}$, Kr$^{5+}$ permitted transitions, while those for Kr$^{6+}$ are taken from \citet{migdalek89}.}
\end{longtable}
}

\end{document}